\documentclass[]{emulateapj}
\usepackage{times, graphicx, setspace, subfigure, latexsym, amssymb, amsmath, fancyhdr, epsf, upgreek}
\usepackage{dcolumn}
\usepackage[backref,breaklinks,colorlinks,citecolor=blue]{hyperref}
\usepackage{natbib}
\usepackage{multirow}
\usepackage{color}
\usepackage{rotating}
\usepackage{subfigure}
\usepackage{cprotect}
\usepackage{afterpage}
\shorttitle{Radio Properties of RRATs}
\shortauthors{B. J. Shapiro-Albert et al.}
\begin{document}

\title{Radio Properties of Rotating Radio Transients: Single-pulse Spectral and Wait-time Analyses}

\email{bjs0024@mix.wvu.edu}

\author{B. J. Shapiro-Albert\altaffilmark{1,2}}
\author{M. A. McLaughlin\altaffilmark{1,2}}
\author{E. F. Keane\altaffilmark{3}}

\altaffiltext{1}{Department of Physics and Astronomy, West Virginia University, Morgantown, WV 26506, USA}
\altaffiltext{2}{Center for Gravitational Waves and Cosmology, West Virginia University, Chestnut Ridge Research Building, Morgantown, WV 26506, USA}
\altaffiltext{3}{SKA Organisation, Jodrell Bank Observatory, SK11 9DL, UK.}


\begin{abstract}
Rotating radio transients (RRATs) are a sub-class of pulsars characterized by sporadic emission and thus can generally only be studied by analysis of their single-pulses. Here we present a single-pulse analysis using 11 years of timing data at 1400~MHz of three RRATs, PSRs~J1819$-$1458, J1317$-$5759, and J1913$+$1330. We perform a spectral analysis on the single-pulses of these RRATs for the first time, finding their mean spectral indices to be  $-1.1 \pm 0.1$, $-0.6 \pm 0.1$, and $-1.2 \pm 0.2$ respectively, within the known range of pulsar spectral indices. We find no evidence for narrowband features as seen for FRB~121102. However, we find the spread of single-pulse spectral indices for these RRATs (ranging from $-7$ to $+4$) to be larger than has been seen in other pulsars, with the exception of the Crab pulsar. We also analyze the time between detected pulses, or wait-time, and find that the pulses are not random and cluster around wait-times of a few pulse periods as well as $\sim 25$ pulse periods for PSRs~J1819$-$1458 and J1317$-$5759. Additionally we find that there is no correlation between the wait-time and pulse flux density. Finally we find that the distribution of the pulse energy for PSRs~J1317$-$5759 and J1913$+$1330 are log-normal, while that of PSR~J1819$-$1458 is log-normal with possible evidence of an additional power-law component.
\end{abstract}


\maketitle
\hypersetup{linkcolor=blue}
\section{Introduction}

Pulsars are useful for many astrophysical studies and efforts to increase the number of known pulsars are ongoing. While most pulsars are discovered in the Fourier domain, single-pulse searches have led to many interesting discoveries. Single-pulse searches of archival Parkes Multibeam data have led to both the discovery of rotating radio transients \citep[RRATs;][]{McLaughlin2006} and fast radio bursts \citep[FRBs;][]{Lorimer2007}. RRATs, a subclass of pulsars, are often not detected in the Fourier domain as their emission is quite sporadic, with only a few detectable pulses per hour\cprotect\footnote{see \verb+http://astro.phys.wvu.edu/rratalog/+} \citep[e.g.][]{Keane2011}, and therefore are preferentially discovered with single-pulse searches. However, since RRAT emission has underlying periodicity, one can time them using techniques similar to those used for normal pulsars \citep{McLaughlin2009}. FRBs are notable for being seen as bright, and highly dispersed pulses of radio emission and thus can only be discovered through single-pulse searches. FRB~121102 is the sole FRB known to repeat, but has no obvious underlying periodicity \citep{Spitler2016}. 

The reason RRATs have such sporadic emission is unknown, but some theories have been put forward to explain this phenomenon. These include fallback of material from a supernova debris disk \citep{Li2006} or interference of an asteroid belt around a pulsar \citep{Cordes2008}. Alternatively, RRATs may be part of the standard canonical pulsar model but exhibit extreme nulling \citep{Wang2007}. 

The sources of FRBs remain an open topic of discussion. FRBs are generally considered to be of extragalactic origin given their larger than expected dispersion measures \citep[DMs; e.g.][]{Bhandari2018}. FRB~121102 has been linked to a host galaxy \citep{Chatterjee2017} confirming its extragalactic origin. Many emission mechanisms have been proposed, including active galactic nucleus emission \citep[e.g.][]{Vieyro2017}, giant flares from magnetars \citep{Lyubarsky2014}, ejection of relativistic high-energy shells from a compact object \citep{Waxman2017}, or the ejection of relativistic jets into surrounding plasma \citep{Romero2016}. Recently, \cite{Michilli2018} found that FRB~121102 has a large rotation measure, indicating the source must be located in an extreme magnetic environment. Additionally, most FRB's also exhibit purely broadband emission, with FRB~121102 also showing narrowband emission \citep{Spitler2016}, and FRB~170827 \citep{Farah2018} showing an interesting narrowband frequency structure. Other FRBs, including FRB~121102, show broadband emission over just part of the band which could be an effect of scintillation \citep{Spitler2018}.

Previous work by \cite{Keane2016} and \cite{Rane2016} has looked into potentially mis-categorizing FRBs as RRATs if they have only been discovered from a single-pulse. While no obvious evidence that any of these single-pulse RRATs should be categorized as an FRB has been found, at DMs near the maximum expected from the Milky Way along a particular line of sight, there is potential for this mis-categorization. Our work does not address this issue as we consider only RRATs that have been observed with many bright pulses within the Galaxy.

We can learn about the emission mechanisms of RRATs by searching for periodicities in pulse arrival times as in \cite{Palliyaguru2011} who found evidence for periodicities in some RRATs on timescales ranging from 1.4 hrs to $\sim5$ yrs. We can also analyze the pulse-amplitude or pulse-energy distributions, as these are useful metrics for comparing different pulsar emission modes. While standard pulsar emission often results in log-normal distributions \citep{BS2012}, so-called ``giant pulses (GRPs)" exhibit power-law distributions and are thought to be due to a different emission mechanism \citep{Cordes2004,Karuppusamy2010}. Additionally, we can look for correlations between the wait-time with the pulse flux density \citep{Cui2017} to examine the emission mechanism.

While most FRBs are single events, FRB~121102 shows large variations in the pulse structure with both frequency and time, exhibiting both broad- and narrow-band emission, with spectral indices ranging from $-10$ to $+14$ at 1400~MHz \citep{Spitler2016}. Similarly, single-pulse spectral index studies on Crab GRPs have found a large spread from $\sim -10$ to $\sim +10$ at 1400~MHz for the main pulse \citep{Karuppusamy2010}. However, while the spread is large, the mean spectral index of the Crab pulsar's main pulse, determined from a Gaussian fit of the spectral index distribution, was found to be $-1.4 \pm 3.3$, where the uncertainty is the standard deviation of the Gaussian. Though not well constrained, this is consistent with previous results from both \cite{Bates2013}, who used pulsar population simulations to find that the mean spectral index for pulsars is $-1.41 \pm 0.06$, and also from \cite{Jankowski2017} who found a mean pulsar spectral index of $-1.60 \pm 0.03$ for 348 pulsars that follow a simple power-law spectrum. In general, almost all pulsars have been found to have negative spectral indices and exhibit broad-band emission \citep[e.g.][]{Lorimer1995, Maron2000, Jankowski2017}. This type of emission is consistent with studies of the Crab, but inconsistent with FRB~121102, which has been observed to have narrow-band emission, and is poorly described by a single spectral index power law \citep{Law2017}.

The standard method of computing the spectral indices of pulsars generally involves fitting a power law to the flux density of the folded profile at multiple different frequency bands across a wide bandwidth \citep[e.g.][]{Lorimer1995, Maron2000, Jankowski2017}. However, folding the pulse profiles loses information, such as pulse-to-pulse profile variation. Similarly if the profile is summed over many frequency subbands then frequency structure can be lost. Similarly, the flux density of the folded profile is often only taken over a single epoch, and thus does not account for any variation on larger timescales.

It has been suggested that, since RRATs are usually analyzed by their single-pulses, they present the opportunity for comparison to FRB~121102. Most RRAT single-pulses have peak flux-density values of $\sim 0.1-1$~Jy \citep{McLaughlin2009}, similar to the peak flux density of many FRBs, which allows us to determine the spectral properties of the single-pulses in a similar way. However, to obtain statistically significant results, we must analyze a large number of single-pulses. Since most RRATs have a low burst rate, many epochs of observation are necessary to observe a large enough sample of pulses. Analyzing data taken over many years has the added benefit of allowing us to account for both single-pulse spectral index and single-pulse energy variation on long timescales.

In the following we analyze 11 years of RRAT timing data, as well as the Parkes Multibeam Pulsar Survey \citep[PMPS;][]{PMSURV2001} data, and conduct a single-pulse and wait-time analysis on three RRATs. We also include a single 7.5 hour observation of PSR~J1819$-$1458 by the Robert C. Byrd Green Bank Telescope (GBT) for additional single-pulse analysis. We describe the different observations in \S \ref{Data}. Our single-pulse search and analysis methods are described in \S \ref{Methods}. We discuss the results of our analyses, presenting the RRAT single-pulse spectral results, wait-time distribution results, and the pulse energy distribution results in \S \ref{Discussion}. Concluding remarks and discussion on future studies are given in \S \ref{Conclusion}.


\section{Data} \label{Data}
The RRAT data come from two main sets, the PMPS \citep{PMSURV2001}, and follow-up observations for RRAT timing taken with Parkes between MJD~52863 (2003 August 12) and MJD~55857 (2011 October 23). An additional 7.5 hour observation of PSR~J1819$-$1458  taken with the GBT is also analyzed.

The PMPS observations were taken with the 13-beam receiver on the 64 m Parkes Radio Telescope between 1998 January and 2002 February. Each observation was 35 minutes in length. The bandwidth for the PMPS data was 288 MHz split into 96, 3 MHz frequency channels centered on 1374 MHz. The data were taken with a sampling rate of 250 $\upmu$s with 1-bit precision \citep{PMSURV2001}.

The majority of the rest of our data are the same as presented in \cite{McLaughlin2009}, but with an additional two years of observations. All observations are between 0.5 and 2 hr long. These data were taken with central beam of the 13-beam receiver on the 64 m Parkes Radio Telescope. Most of the observations were taken with a central frequency of 1390 MHz, a bandwidth of 256 MHz and 512 frequency channels, and were taken with a sampling rate of 100 $\upmu$s with 1-bit precision.
A minority of RRAT timing observations were taken with the $10-50$ cm receiver, which has a bandwidth of 64 MHz around 685 MHz and a bandwidth of 768 MHz around 3 GHz, and the HOH receiver, which has a bandwidth of 576 MHz and a central frequency of 1500 MHz. For consistency in our data set we have ignored these data in our analysis.

In addition to the data from Parkes, we have also separately analyzed a 7.5 hr observation of PSR~J1819$-$1458 taken with the GBT on MJD~54557 (2008 April 1). This observation was taken with a bandwidth of 800 MHz and a center frequency of 2 GHz. These data were taken at a sampling rate of 81.92~$\upmu\rm{s}$ with 8-bit precision. However we have only the dedispersed, frequency-scrunched, time series for this observation and thus have no spectral information.

\section{Methods} \label{Methods}

We first describe the methods used to determine the single-pulse spectral indices. We then analyze simulated data using these same methods to determine if an injected spectral index could be recovered. We then determine which RRATs are viable candidates for our analysis.

Additionally, we analyze the distributions of pulse wait-times. If the emission on short timescales is purely random, these distributions will be exponential. However, deviation from an exponential suggests that there may be some periodicities to the emission on short timescales. We test this by fitting a variety of distributions in various combinations and performing multiple goodness-of-fit tests on these.

Finally we look for correlations in the flux density of the single pulses with wait time. An increase in flux density with wait-time could suggest that the emission is due to a build-up of energy in the pulsar magnetosphere.

\subsection{Single-pulse Spectral Index Analysis} \label{SinglePulseMethod}
\subsubsection{Identification of Single pulses} \label{SPID}

To identify single pulses from RRATs, we use the same method as described in \cite{McLaughlin2009}. Using \verb+SEEK+ and other packages in \verb+SIGPROC+\cprotect\footnote{\verb+http://sigproc.sourceforge.net/+} \citep{Lorimer2011} to search for pulses at a $5\sigma$ threshold, the data are dedispersed with DM of both 0~pc~cm$^{-3}$ and the known DM of the RRAT (see Table \ref{RRATParams}). \verb+SEEK+ may underestimate the signal-to-noise (S/N) of a single-pulse \citep{Keane2015}. However the S/N returned by \verb+SEEK+ is used just for initial thresholding purposes, and the actual S/N of the pulse is calculated during the fitting and calibration steps of our analysis. It is possible that we may not detect some weak pulses, but we estimate that even in a worst-case scenario where, assuming the pulses are randomly emitted, a pulse will fall into every underestimated phase of the \verb+SEEK+ boxcar, we will only miss only one out of every five pulses detected with an S/N of 7$\sigma$ or less. This is at most $\sim3$\% of the pulses found for a RRAT with a $\sim 1$~s period and a pulse width of $\sim 10$~ms. As the goal of this work requires only a large population of bright pulses, this should not significantly impact our results.

Pulse candidates at both DMs are then compared and any pulses detected with a higher S/N at a DM of 0~pc~cm$^{-3}$ are discarded as radio frequency interference (RFI), as pulses from the RRAT will be brighter at the true DM. We set a minimum detection threshold of $5\sigma$ for each pulse.

As an additional guarantee that we have filtered out all RFI, we only take pulses whose times of arrival (TOAs) based on the brightest pulse in an observation occur within 5\% of the expected phase. For the purposes of our wait-time analysis, we then round the wait-time such that it is an integer number of pulse periods. After this filtering, each pulse is also visually inspected and any remaining RFI is manually discarded.

\begin{figure}[!t]
\includegraphics[width = 9cm]{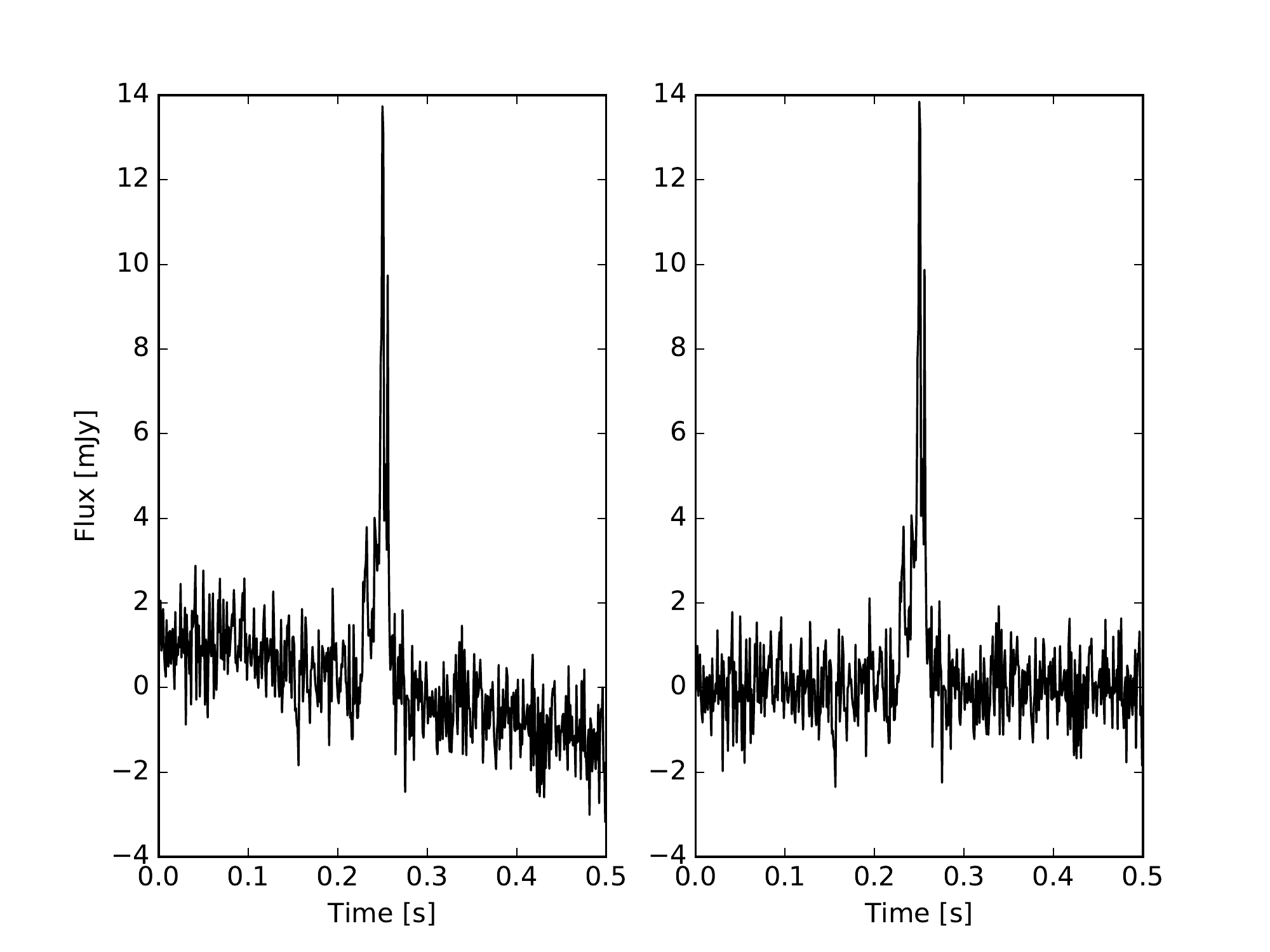}
\caption{Single pulse from the GBT observation of PSR~J1819$-$1458. Left: the pulse after calibration, but before the Gaussian Process Regression (GPR) has been implemented. We see that the baseline is obviously sloped. Right: the pulse post-GPR. We see that the slope of the baseline has been eliminated without compromising the pulse profile. \label{GPREx}}
\end{figure}

\subsubsection{Fitting Single-pulse Templates} \label{SPTempfits}

After the pulses have been filtered using the methods described above, we take 0.5 s of dedispersed data split into 512 time bins around each pulse. This is done to reduce the amount of time necessary to process the data. From simulations (see \S \ref{FakeAnalysis}) we find the optimal number of subbands for each set of observations. Each single-pulse from the timing data of \cite{McLaughlin2009} is split into 16 subbands. When dedispersing the PMPS observations \citep{PMSURV2001}, each pulse is split into 12 subbands. Since we must be able to detect the pulse in each subband, we take only pulses detected above an S/N threshold of $8 \sigma$. In order to determine the flux density of each pulse, we calibrate the data using the radiometer equation \citep{HANDBOOK},
\begin{equation} \label{radiometer}
\Delta S_{\rm{sys}} = \dfrac{\beta T_{\rm{sys}}}{G \sqrt{n_{\rm{p}} t_{\rm{obs}} \Delta f}} = C \sigma_{\rm{p}},
\end{equation}
where our system temperature, $T_{\rm{sys}}$, is the receiver temperature, $T_{\rm{rec}}$, plus the sky temperature, $T_{\rm{sky}}$. $T_{\rm{sky}}$ is determined at the center frequency of each subband scaling from the 408 MHz all sky survey of \cite{Haslam1981} assuming a spectral index of $-2.6$ \citep{Lawson1987}. For the Parkes Multibeam Receiver, $T_{\rm{rec}}$ is taken to be 28 K (2017 October 14 version of the ``Parkes Radio Telescope Users Guide"). For the GBT S-band receiver, $T_{\rm{rec}}$ is taken to be 20 K (2017 February 14 version of ``Observing With The GBT"). For both Parkes and the GBT, spillover and atmospheric contributions to $T_{\rm{sys}}$ are negligible at the frequencies considered here. The $\beta$ factor accounts for loss due to 1-bit digitization and is $\sqrt{\pi/2}$ for the Parkes observations, and $\sim 1$ for the GBT observation as it records with 8-bit precision. Our data are multiplied by the resulting conversion factor, $C$, in order to convert arbitrary units of flux density to mJy. Finally, $\sigma_{\rm{p}}$ is the standard deviation of the off-pulse region.

We then remove any variations in the baseline that occur due to RFI or instrumental effects. Most of our data were not badly affected by these variations, but, as shown in the left panel of Figure \ref{GPREx}, removal of these trends was necessary in some cases. We removed these variations by Gaussian Process Regression \citep[GPR;][]{GPRBook} which is implemented through the \verb+python+ package \verb+GPy+\cprotect\footnote{\verb+https://sheffieldml.github.io/GPy/+}. We fit the baseline only in the off-pulse region, determined as the area outside of twice the full width at half max (FWHM) of the pulse measured from its peak. This is visually checked to make sure that no part of the pulse in within this off-pulse region. Only variations more than 1$\sigma$ away from a zero mean with length scales of at least $\sim$0.05 s are considered. The fit over the baseline is then subtracted from the pulse to remove these variations. An example of a single-pulse after GPR is shown in the right panel of Figure \ref{GPREx}.

Once the pulse has been calibrated and undergone GPR, we fit a template profile to the pulse using a least-squares minimization technique. Performing GPR on each pulse before the fitting is necessary because the pulse in each subband is usually noisy and has little flux density, so having a prior template that is immune to this noise allows for better subband fitting. We do not subtract this baseline from each subband, as we fit only the amplitudes of each component of our template in each subband. To do this fitting, we use the \verb+leastsq+ function in the \verb+python+ package \verb+scipy+\cprotect\footnote{\verb+https://www.scipy.org/+} \citep{Jones2001}. 

One of the most well studied RRATs, PSR~J1819$-$1458, has three components in its profile \citep{Karastergiou2009}, so we allow our profile template to fit up to three Gaussian components to each single-pulse. This also helps to account for variable small-scale structure in the pulse profile. Our total profile integrated over frequency is fit by
\begin{equation} \label{pulseprof}
P(t) = \sum^{3}_{i = 1} A_{i}\exp{ \left ( -\frac{(t-t_{p_{i}})^{2}}{2\sigma_{i}^{2}} \right ) },
\end{equation}
where $t$ is the time of the pulse, $i$ the $i$th pulse component, and $A$, the amplitude, $t_{p}$, the time of the pulse peak, and $\sigma$, the width, are the free parameters of the pulse profile. After the first component is fit, we subtract the fit from the data and a second component is fit from the residual. The three-component fit is made by subtracting the two-component fit from the data and then fitting a third component from the residual. The third component is fit even if no significant secondary component is found. If the pulse does not have a significant second or third component, the Gaussian component that is fit will have zero amplitude.

This fitting method estimates an error matrix for the template using the partial derivatives of the Gaussian components. Since some of our pulses are weak, in order to add a second or third Gaussian component to the template we require the reduced chi-squared value, $\chi^{2}_{r}$, to be at least $10\%$ better than a fit with fewer components.

The standard deviation used for the $\chi^{2}_{r}$ of each template is taken as the root mean square of the off-pulse around that single-pulse. We define the on-pulse region as twice the FWHM of the single Gaussian fit determined by fitting a single Gaussian component to the pulse using the least-squares fitting method described above. The region outside of this is the off-pulse. 

\subsubsection{Obtaining a Single-pulse Spectral Index} \label{SPalpha}

To fit a spectral index to each single pulse, we fit the amplitude, or flux density, of the pulse in each subband. We assume that the number of components, the phase, and the width of each component do not change between the subbands and our composite fit. We also assume that we have averaged over many scintles in each subband (see \S \ref{RRATsUsed}) so our analysis is not limited by scintillation. We then perform the same calibration described above on each subband. Our least-squares fitting in the subband fits only for the amplitude of each component. The uncertainty in this fit is again taken from the error matrix and is based on the partial derivatives. 

However, the partial derivatives of a Gaussian increase dramatically if components are either very narrow or very close to each other. This is due to the error inherent in our template fitting, and is not physical. We check for this by testing if the fitting error on the pulse amplitude is greater than 1000\%, as this is where this issue manifests most clearly. If the partial derivatives meet this criterion, we account for it by assuming a 50\% template fitting uncertainty. This down-weights the flux density in the subband so that it has smaller impact on our fit spectral index. 

We then integrate the Gaussian template of the pulse over the half second of data we record to obtain an estimate of the flux density in each subband,
\begin{equation} \label{GIntegral}
S(A, \delta, \sigma) = \int_{0}^{0.5} P(t)~{\rm{d}}t,
\end{equation}
where $P(t)$ is defined as in Eq. \ref{pulseprof}. We note that, while integrating the template minimizes the amount of noise included in the flux density, it is likely to underestimate the true flux density of the pulse, as small pulse structures may not meet our $\chi^{2}_r$ constraint. 

To calculate the uncertainty of the subband flux density, we take
\begin{equation} \label{GIntPartial}
\sigma_{S} = \left [\sum^{3}_{i =1} 2 \pi (\sigma_{i}^{2} \sigma_{A_{i}}^{2} + A_{i}^{2} \sigma_{w_{i}}^{2}) \right ]^{\frac{1}{2}},
\end{equation}
where $A$, $\sigma$, are defined as above, $\sigma_{A_{i}}$ is the uncertainty in the amplitude, and $\sigma_{w_{i}}$ the uncertainty in the width. We do this for each subband of the pulse and then use a weighted least-squares fit to determine the spectral index of the pulse. 

\begin{figure}
\centering
\includegraphics[width=\columnwidth]{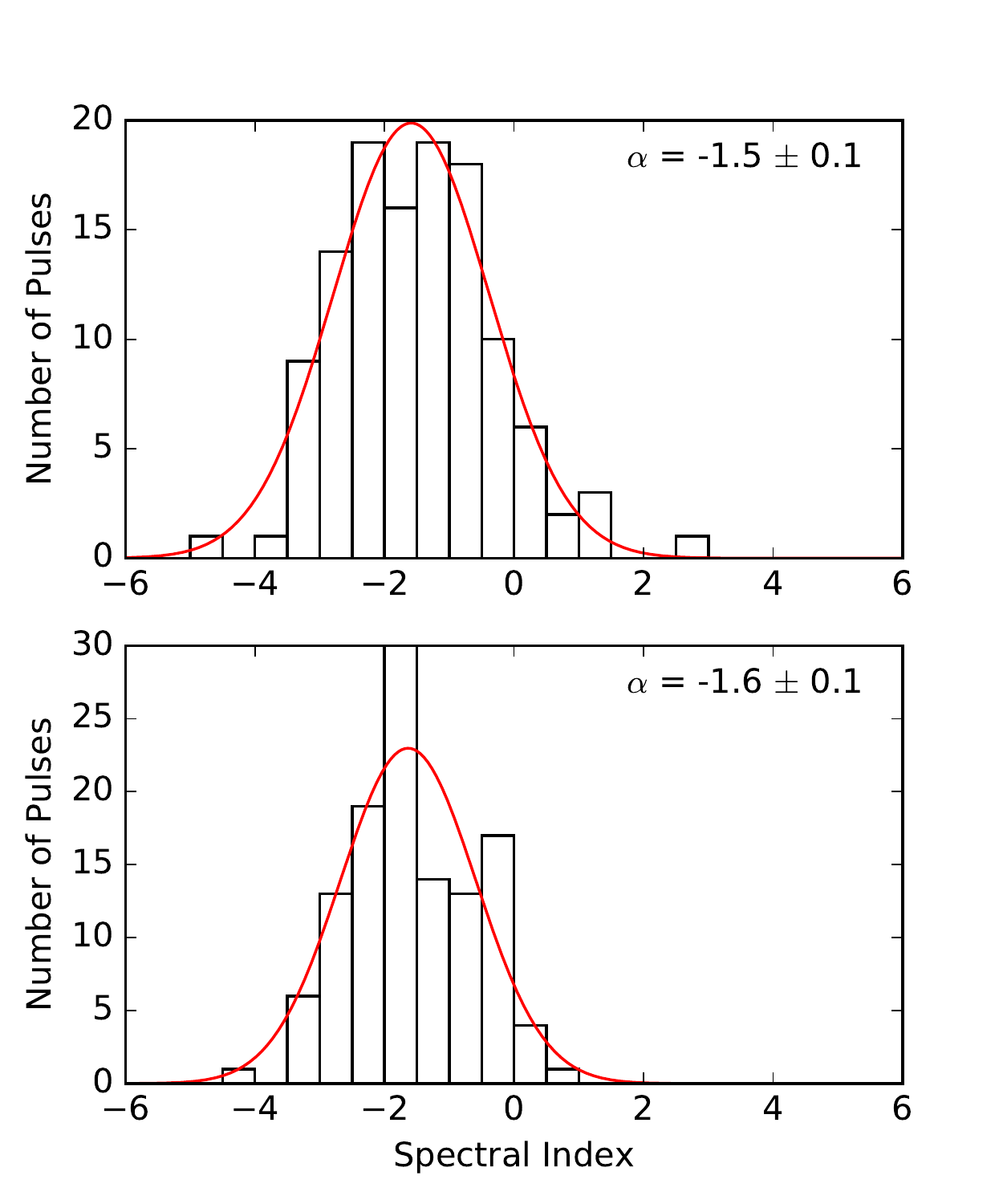}
\caption{Top: distribution of the spectral indices for the simulated PMPS-style data with an input spectral index of $-1.5$. The weighted mean spectral index with mean errors is reported in the upper right corner as $\alpha = -1.5 \pm 0.1$ and a Gaussian fit is shown over the histogram. The standard deviation of the Gaussian is $1.2$. Bottom: distribution of the spectral indices for the simulated RRAT timing-style data with an input spectral index of $-1.5$. The weighted mean spectral index with mean errors is $\alpha = -1.6 \pm 0.1$. The standard deviation of the Gaussian is $1.0$. The timing data are more precise since they have more frequency subbands, which accounts for the slightly narrower Gaussian fit to the simulated data. \label{FakeHists}}
\end{figure}

\begin{table*}[h!t]
\begin{center}
\caption{RRAT Parameters \label{RRATParams}}
\begin{tabular}{c c c c c c c c}
\tableline
Name & Period & DM & R.A. (J2000) & Decl. (J2000) & $T_{\rm{obs}}$ & $N_{5\sigma}$ & Pulse Rate \\
 & (s) & (pc cm$^{-3}$) & (hms) & ($^{\circ}$ $' "$ ) & (hr) & (pulses) & (h$ ^{-1} $) \\
\tableline
PSR J1819$-$1458 & 4.263 & 196(3) & 18:19:33.0(5) & $–$14:58:16(32) &  44.8 & 1170 (Parkes) & $\sim 26$ \\
 & & & & & 7.5 & 937 (GBT) & $\sim 125$ \\
PSR J1913$+$1330 & 0.923 & 175.64(6) & 19:13:17.975(8) & +13:30:32.8(1) & 28.9 & 228 & $\sim 8$\\
PSR J1317$-$5759 & 2.642 & 145.3(3) & 13:17:46.29(3) & $−$57:59:30.5(3)  & 61.3 & 289 & $\sim 5$ \\
\tableline
\end{tabular}
\end{center}
\tablecomments{Parameters for RRATs analyzed in this work. All parameters are from \cite{McLaughlin2009} except those for PSR~J1819$-$1458 which are from \cite{Lyne2009}. $T_{\rm{obs}}$ is the total time spent on each object and $N_{5\sigma}$ is the total number of pulses found above a 5$\sigma$ detection threshold.}
\end{table*}

We assume the flux density varies as a power law,
\begin{equation} \label{SpecIdxEq}
S \propto \nu^{\alpha},
\end{equation}
where $S$ is the total integrated flux density of our pulse template in the subband, $\nu$ the center frequency of the subband, and $\alpha$ the spectral index of the pulse. If the amplitude of a Gaussian component of the pulse profile fit is returned as a negative value, it is set to zero. We remove from the spectral index fit any subband where the computed flux density is more than $2\sigma$ below the mean flux density of the other subbands.

While removing these points could bias our fit, our weighted fit will weight subbands with the largest error bars the least. Therefore the removal of these subbands will have a minimal effect on the fit spectral index.  

The uncertainty in the fit spectral index of the each single-pulse is given from the covariance matrix produced by the least-squares fit as described above. The mean spectral index is determined by computing the weighted mean of the spectral indices of all of the single-pulses.

\subsection{Analysis of Simulated Data} \label{FakeAnalysis}

In order to verify that the methods described in \S \ref{SinglePulseMethod} return the correct spectral indices, we simulated data from both the PMPS as well as the RRAT timing observations using the \verb+fake+ function available in \verb+SIGPROC+-v4.3 \citep{Lorimer2011}. The simulated PMPS observation has the same parameters as described in \S \ref{Data} and was given a length of 2~min, a period of 1~s, pulse width of 0.01~s, and an injected single-pulse peak S/N of 2. These parameters allowed each pulse to be detected by \verb+SEEK+ with a maximum S/N above our threshold of $8\sigma$. As noted earlier, \verb+SEEK+ can also underestimate the S/N of a pulse based on the pulse phase of the boxcar searching algorithm \citep{Keane2015}. This variation, in addition to the variation due to smoothing the data, returned all 120 pulses with an S/N between 9 and 12$\sigma$. We then multiplied each frequency channel by an appropriate scaling factor to inject the simulated data with a spectral index of $-1.5$.

After simulation and detection of the single-pulses, we then used the method described in \S \ref{SinglePulseMethod} to fit a spectral index to each pulse and calculate the mean spectral index. When calibrating the simulated data, we used Eq. \ref{radiometer} but took $T_{\rm{sys}}$=$T_{\rm{rec}}$ because \verb+fake+ does not simulate the sky temperature and since $T_{\rm{sky}} \propto \nu^{-2.6}$ \citep{Lawson1987}, adding this induces a separate spectral index not initially present in our simulated data. 

In order to determine if the number of subbands used is significant, the method was repeated with the data split into 6, 12, 24, 48, and 96 subbands. Depending on the steepness of the spectral index of a single-pulse, across any given subband, the spectral index may not be flat. However, for a spectral index of $-1.5$, the largest subband tested was 48 MHz wide, which exhibits a change in the spectral index of 0.03 (e.g. from $-1.5$ to $-1.53$) across one subband. This is much smaller than the uncertainty on any given spectral index as well as on the weighted mean spectral index for the entire distribution. We therefore assume that the spectral index across any given subband is flat.

For each case, the weighted mean spectral index was calculated and a Gaussian was fit with a least-squares method to the distribution of spectral indices. In all cases, the mean spectral index of the pulses was recovered at $-1.5$ within 1$\sigma$. The reduced chi-squared value for each Gaussian fit was also determined. Finally, we used our analysis software (see \S \ref{SinglePulseMethod}) to determine the actual S/N distribution and analytically determined what the distribution of spectral indices should be based on this. Based on the criteria described above, we determined that using 12 subbands accurately recovered our injected spectral index of $-1.5$ and using more subbands did not improve our recovered values. The distribution of spectral indices for this set of simulate data using 12 subbands is shown in the top panel of Figure \ref{FakeHists}.

\begin{figure*}[p]
\centering
\includegraphics[width=\textwidth]{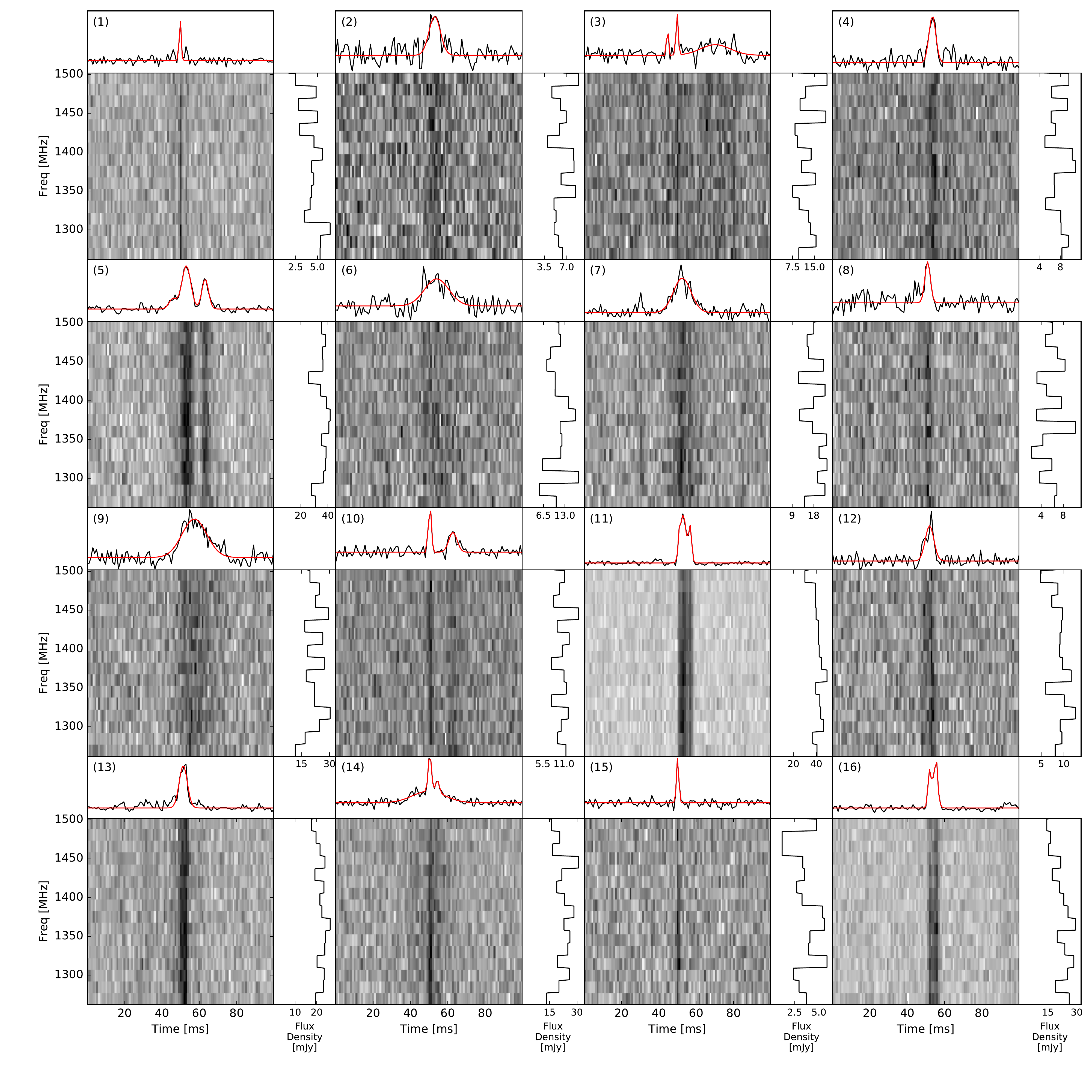}
\caption{Sample of single pulses from PSR~J1819$-$1458 from a single 30 minute Parkes timing observation. The top panel of each plot shows the frequency-summed composite pulse in black, with the Gaussian template fit in red. The dedispersed frequency-time plot for the pulse is shown below the composite pulse panel. To the right is a measure of the mean flux density in mJy of the pulse in each subband. Time and frequency axes are shared by all pulses in the same column and row respectively. This sample illustrates the wide variety of single-pulses we observe. In Figure \ref{SpecIdx} we show the spectral fits for these pulses. \label{BrightPulses}}
\end{figure*}

\begin{figure*}
\includegraphics[width=\textwidth]{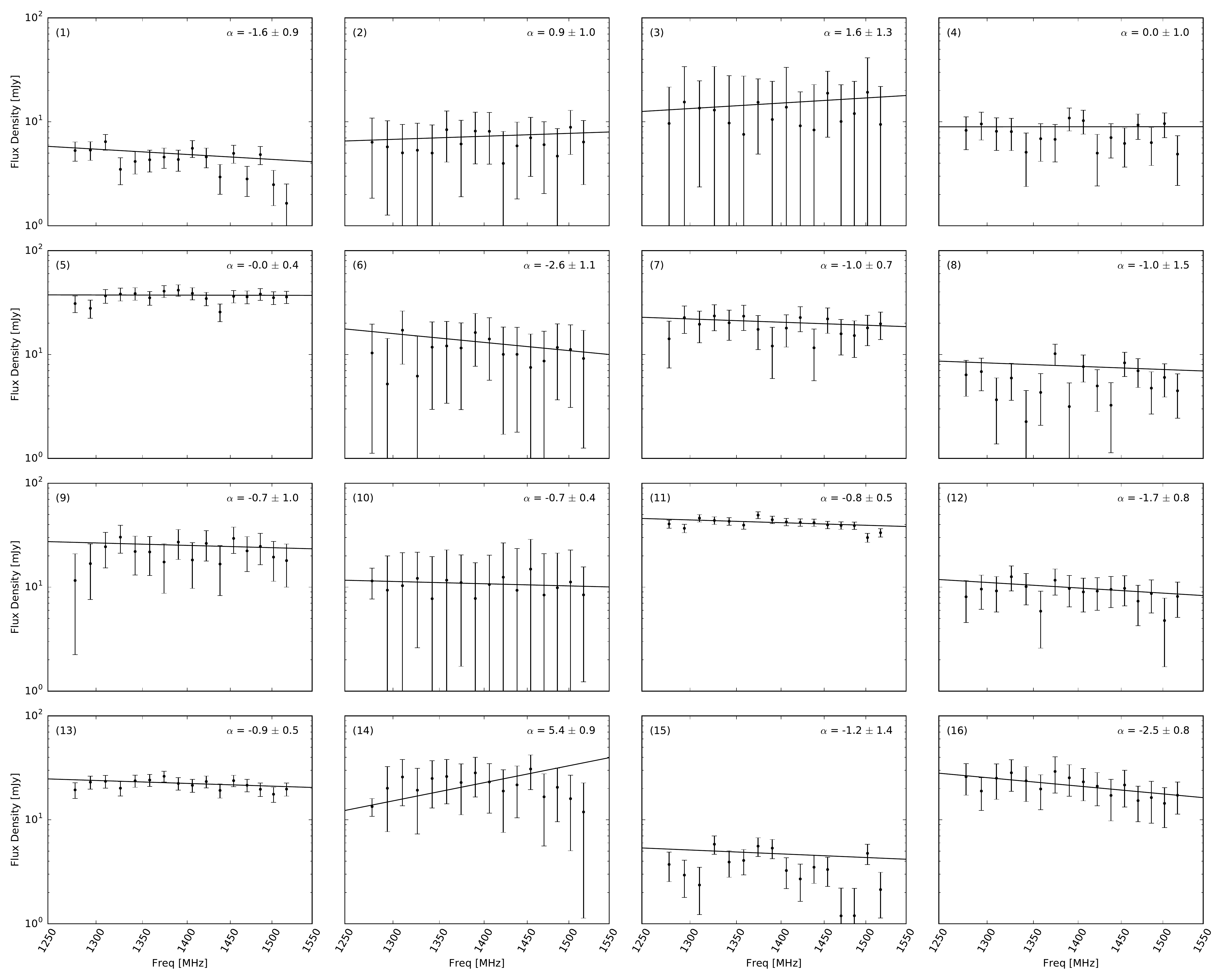}
\caption{Sample of single-pulse spectral index fits from  PSR~J1819$-$1458. Each single-pulse spectral index fit can be referenced to the corresponding single-pulse in Figure \ref{BrightPulses}. The frequency and flux density scales are the same for each pulse spectral index fit. The scaled mean flux density from Equation \ref{SpecIdxEq} is plotted against the frequency in log space. \label{SpecIdx}}
\end{figure*}

\begin{figure}[ht]
\centering
\includegraphics[width=\columnwidth]{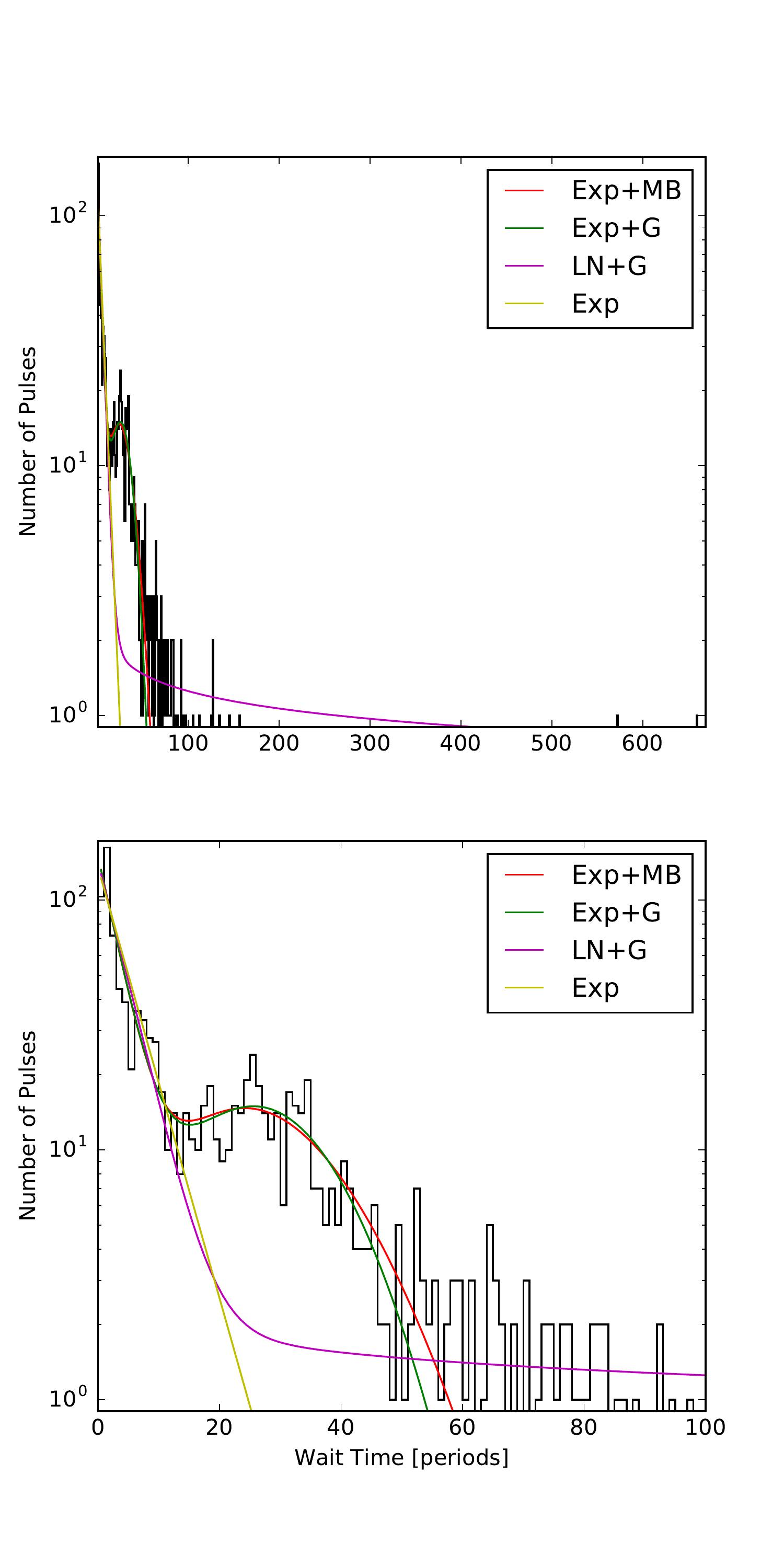}
\caption{Upper panel: distributions of single-pulse wait-times for the Parkes observations of PSR~J1819$-$1458 at L-band. We see that most pulses come within a couple of periods of each other. We also see that there is an extended tail of long wait-times and a small secondary distribution around 25 periods. We fit the full distribution with an exponential plus a Gaussian (Exp+G; green), an exponential plus a Maxwell-Boltzmann distribution (Exp+MB; red), a log-normal plus a Gaussian (LN+G; magenta), and a pure exponential (Exp; yellow). Lower panel: same as the upper panel but zoomed in on the first 100 periods. \label{J1819ParkesWTHist}}
\end{figure}

To simulate the later RRAT timing observations, we use the same observation parameters as in \S \ref{Data} and the same pulsar parameters as with the simulated PMPS data. For this set, 120 single-pulses were found above our 8$\sigma$ threshold with most having a maximum S/N value reported by \verb+SEEK+ between 12 and 16$\sigma$. The difference in detection significance between this set of simulated data and the simulated PMPS data is due to the method by which the boxcar used by \verb+SEEK+ smooths the data. To check if the number of subbands used is significant, our method was tested on this simulated data set using 4, 8, 16, 32, 64, and 128 subbands. We used the same criteria described for the simulated PMPS data to determine which number of subbands best recovered our initial values. We found that 16 subbands most closely recovered the injected spectral index, as shown in the bottom panel of Figure \ref{FakeHists}.

The difference in the optimal number of subbands is due to the fact that the even divisors of the total number of subbands (96 for the PMPS data and 512 for the RRAT timing observations) are different. For the rest of our spectral index analysis, we have split the PMPS pulses into 12 subbands, and the pulses from all other observations into 16 subbands.

\begin{table}[b]
\begin{center}
\caption{RRAT Scattering Parameters \label{RRATscatparams}}
\begin{tabular}{c c c c c c c c}
\tableline
     & \multicolumn{4}{c}{NE2001} & \multicolumn{3}{c}{YMW16} \\
Name & $\Delta \nu_{\rm{d}}$ & $\Delta t_{\rm{d}}$ & $\tau_{\rm{d}}$ & $\Delta \tau_{\rm{d}}$ & $\Delta \nu_{\rm{d}}$ & $\Delta t_{\rm{d}}$ & $\tau_{\rm{d}}$ \\
     & (kHz) & (s) & ($\upmu$s) & ($\upmu$s) & (kHz) & (s) & ($\upmu$s) \\
\tableline
PSR~J$1819-1458$ & 1.3 & 13 & 80 & 70 & 0.5 & 7 & 30 \\
PSR~J$1913+1330$ & 4.8 & 31 & 20 & 10 & 1.6 & 14 & 20 \\
PSR~J$1317-5759$ & 30 & 52 & 4.3 & 4 & 0.7 & 12 & 1 \\
\tableline
\end{tabular}
\end{center}
\tablecomments{Scintillation and scattering parameters for the three RRATs from both NE2001 \citep{NE2001} and YMW16 \citep{YMW2017} DM models. We note that $\Delta \nu_{\rm{d}}$, $\Delta t_{\rm{d}}$, and $\tau_{\rm{d}}$ are taken at the center of our band. However, $\Delta \tau_{\rm{d}}$ for NE2001 denotes the difference in pulse width due to scattering between the top of the band and the bottom, or the change in $\tau_{\rm{d}}$ across our band. The YMW16 $\tau_{\rm{d}}$ denotes the scattering at just the center of our band. We do not calculate $\Delta \tau_{\rm{d}}$ for the YMW16 model since their values of $\tau_{\rm{d}}$ come purely from a scaling law in \cite{Bhat2004} based on DM. }
\end{table}

\subsection{RRATs Analyzed in This Work} \label{RRATsUsed}

After initial testing using the methods described in \S \ref{SinglePulseMethod}, our work focuses on PSRs~J1819$-$1458, J1913$+$1330, and J1317$-$5759 as no other RRATs in our data set had enough detectable bright pulses for a complete analysis \citep{McLaughlin2009}. Useful parameters for these RRATs can be seen in Table \ref{RRATParams}. We note that sky position and spectral index are covariant when the uncertainties on the former are large; however the uncertainty on the positions of these three sources is such that this covariance is broken (see Table \ref{RRATParams}).

For these three RRATs, the predicted diffractive scintillation timescales, $\Delta t_{\rm{d}}$, and bandwidths, $\Delta \nu_{\rm{d}}$, as well as the scattering timescale $\tau_{\rm{d}}$, have been calculated using the NE2001 model described in \cite{NE2001}, assuming a source velocity of 100 km~s$^{-1}$ at a center frequency of 1390~MHz. Additionally, for the NE2001 model we have calculated $\Delta \tau_{\rm{d}}$, the difference in $\tau_{\rm{d}}$ between the top and the bottom of the band. This allows us to see how much the pulse width changes between the top and bottom of the band due to scattering and check that our assumption that the pulse is the same width in each subband holds. These values are reported in Table \ref{RRATscatparams}.

While the YMW16 electron density model \citep{YMW2017} does not explicitly estimate scintillation parameters, one can estimate the scintillation bandwidth from
\begin{equation} \label{scintband}
2 \pi \Delta \nu_{\rm{d}} \tau_{\rm{d}} = C_{1},
\end{equation}
where $\tau_{\rm{d}}$ is the scattering timescale returned from the YMW16 model at the center frequency of the band estimated using the DM scaling equations from \cite{Bhat2004}. We take $C_{1} = 1$ for a thin scattering screen. We then estimate the scintillation timescale using
\begin{equation} \label{scinttime}
\Delta t_{\rm{d}} = A_{\rm{ISS}} \frac{\sqrt{D \Delta \nu_{\rm{d}}}}{V_{\rm{ISS}} \nu}.
\end{equation}
from \cite{Cordes1998}, where $D$ is the distance in kpc, $\Delta \nu_{\rm{d}}$ is in MHz, $\nu$ is the observing frequency in GHz, $V_{\rm{ISS}}$ the velocity of the pulsar assumed to be 100~km~s$^{-1}$, and $A_{\rm{ISS}} = 2.53 \times 10^{4}$~km~s$^{-1}$ for a uniform medium. The estimated values of $\Delta t_{\rm{d}}$ and $\Delta \nu_{\rm{d}}$ are reported in Table \ref{RRATscatparams}. Additionally we report the values of $\tau_{\rm{d}}$ but we do not report the change in scattering width, $\Delta \tau_{\rm{d}}$, across the band from the YMW16 model as $\tau_{\rm{d}}$ comes purely from the \cite{Bhat2004} scaling laws and is not modeled separately as in NE2001.

While $\Delta t_{\rm{d}}$ are on the same order as the period of these RRATs, $\Delta \nu_{\rm{d}}$ from both NE2001 and YMW16 are all $< 1$~MHz, so for subbands of 16~MHz or greater, we are likely to average over many scintles and can therefore neglect the effects of scintillation in our analysis. Additionally, the difference in $\tau_{\rm{d}}$ between the top and bottom of our band, $\Delta \tau_{\rm{d}}$, is smaller than our time resolution, and confirms our assumption that the width of the pulse does not change substantially between subbands. 

Specific to PSR~J1819$-$1458 is its three-component profile. The components of this profile are known to vary by $\pm \sim 45$~ms of the expected pulse phase \citep{Lyne2009}. Since we reference our TOAs to the brightest pulse in the observation and we do not know which part of the profile our detection is from, we use a TOA window of 6\% of the expected pulse phase (instead of 5\%) in order to account for all pulses in the observation.

A sample of the single-pulses analyzed and their template fits can be see in Figure \ref{BrightPulses}. While the sample shown is small compared to the total number of single-pulses detected, the wide variety we detect is apparent. While it can be seen that our template fitting may not fit components that appear to be real (e.g. pulses 1 or 7), this is due to the $>10\%$ improvement in $\chi^{2}_{r}$ that we require to add a component, and prevents us from fitting noise into our pulse template as may occur in noisier pulses such as pulses 2 or 4. The spectral index fits for these single-pulses are shown in Figure \ref{SpecIdx}.

\begin{figure*}
\includegraphics[width = \textwidth, height=5cm]{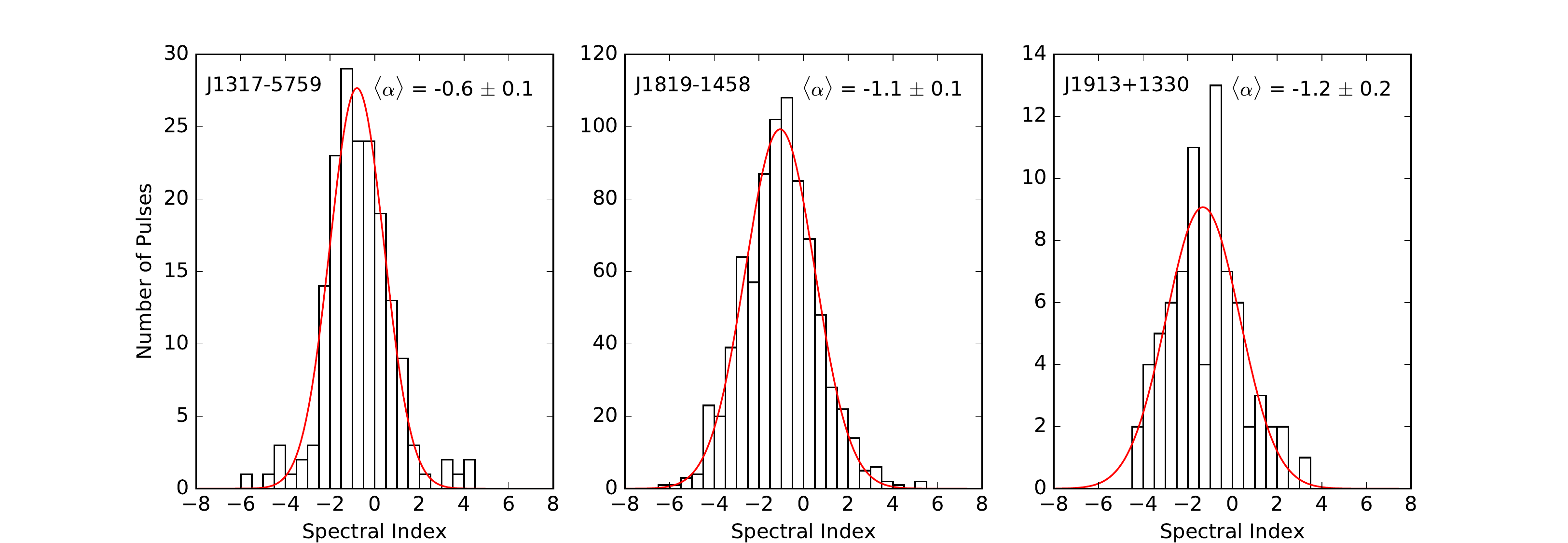}
\caption{Distribution of single-pulse spectral indices for the RRATs analyzed in this work. The RRAT corresponding to each distribution is shown in the upper left corner, and the mean spectral index of each RRAT is reported in the upper right corner. The error reported on the spectral index is the error on the mean value. \label{AlphaHists}}
\end{figure*}

\subsection{Wait-time Analysis} \label{WTmethod}

We use the same single-pulses found using the methods in \S \ref{SinglePulseMethod} in our wait-time analysis; however, since subband sensitivity is not required, all pulses found at the $5\sigma$ threshold are used. The wait-time for each pulse is determined by calculating the time between confirmed astrophysical pulses within an observation. The standard Parkes observation length is $\sim 422$~pulse periods for J1819$-$1458, $\sim 1950$~pulse periods for J1913$+$1330, and $\sim 681$~pulse periods for J1317$-$5759, with a few observations up to four times as long for each RRAT. Since we do not know how many pulses may have occurred between observations, no wait-time is reported between observations, so the longest possible wait-time is $\sim 2$~hr. 

In order to analyze the wait-time$-$flux-density relation for each RRAT, we bin the pulse wait-times by period and can then calculate the weighted mean flux density in each bin. We do not include any pulses where the error on the peak flux-density amplitude was larger than the calculated peak flux-density amplitude.

If the RRAT emission was described by a uniform distribution in time, we would expect the distribution of our pulse wait-times to be exponential. As a motivating example, the distribution of wait-time for the Parkes observations of PSR~J1819$-$1458 is shown in Figure \ref{J1819ParkesWTHist}. We see that most pulses seem to be in groups, with the majority of pulses coming within a few periods of each other. There are also distinct tails to these distributions that suggest the distribution is not exponential. 

We fit these distributions using the same least-squares fitting algorithm as in \S \ref{SinglePulseMethod} using four models. We first fit a pure exponential function and then fit more complex models to the wait-time distributions, including an exponential plus a Gaussian, an exponential plus a Maxwell-Boltzmann, and a log-normal plus a Gaussian. All models can be seen in Figure \ref{J1819ParkesWTHist}.

\subsection{Pulse Energy Analysis} \label{Ampmethod}

After calibrating each single-pulse found in \S \ref{SinglePulseMethod} using Eq. \ref{radiometer}, we use the total flux density of each pulse, or the pulse energy, found using the same methods and filtering as in our wait-time analysis, to analyze the pulse energy distribution. The pulse energy is not dependent on bin size and is thus a more robust way of characterizing the emission mechanism than peak flux. A similar analysis on pulse amplitude distributions by \cite{Cui2017} found that most RRATs have a log-normal pulse amplitude distribution with little evidence of power-law tails, which describe GRP amplitude distributions \citep{Mickaliger2012}. 

We use a similar method to \cite{Cui2017} to fit our pulse energy distributions for our three RRATs using three different models. We use a pure power-law model, a pure log-normal distribution, and finally a combined power-law and log-normal distribution. When fitting these models to our power-law distributions, we fit only from the pulse energy at which $N(S_{p})$ is largest up to higher energies, as we cannot fit the energy distributions of weak pulses well with a log-normal, power-law, or combined model.

\section{Results and Discussion} \label{Discussion}

Our single-pulse analysis of the RRATs PSRs~J1819$-$1458, J1913$+$1330, and J1317$-$5759 finds that the spectral indices of the single-pulses for all three RRATs follow a normal distribution. This is in line with work by both \cite{Kramer2003}, who analyzed the single-pulse spectral indices of PSRs~B0329$+$54 and B1133$+$16, and \cite{Karuppusamy2010}, who analyzed Crab GRP spectra. Both studies found the single-pulse spectral index distributions well represented by a normal distribution. 

Our analysis has also shown that the pulses do not seem to be emitted randomly, and we discuss the implications of this for RRAT emission mechanisms. Additionally we find that the time between pulses is not correlated with pulse flux density. Finally, we find that the distributions of the pulse energy for PSRs~J1317$-$5759 and J1913$+$1330 are log-normal, while that of PSR~J1819$-$1458 is log-normal with possible evidence of an additional power-law component.

\subsection{Spectral Indices} \label{SpecIdxDisc}
In this work we have analyzed the spectral indices of single-pulses of RRATs for the first time. We find that, while there is a wide spread of spectral indices, they are well described by a Gaussian distribution. The peaks of the Gaussian distributions agree within $1 \sigma$ with both the weighted and unweighted mean spectral indices for each respective RRAT. The distributions for all the three RRATs can be seen in Figure \ref{AlphaHists}. We also report the mean spectral index and the number of single-pulses found for each RRAT in Table \ref{RRATspecResults}.

We note that there is a large amount of pulse-to-pulse variability in the measured single-pulse spectral indices shown in Figure \ref{AlphaHists}, with a difference of $\sim$10 between the highest and lowest values we find. There are very few other single-pulse spectral index studies in the literature for a comparison of our results. One single-pulse study of two pulsars, PSRs~B0329$+$54 and B1133$+$16, by \cite{Kramer2003} found that the difference between the minimum and maximum spectral indices was $\sim 3$. However they also found that the individual components of the single-pulses spanned a slight larger range of spectral indices of $\sim 4$. 

\begin{table} [t]
\begin{center}
\caption{RRAT Mean Spectral Indices \label{RRATspecResults}}
\begin{tabular}{c c c}
\tableline
Name & $N_{8\sigma}$ & $\langle \alpha \rangle$ \\
 & (pulses) & \\
\tableline
PSR~J$1819-1458$ & 797 & $-1.1 \pm 0.1$ \\
PSR~J$1913+1330$ & 75 & $-1.2 \pm 0.2$ \\
PSR~J$1317-5759$ & 171 & $-0.6 \pm 0.1$ \\
\tableline
\end{tabular}
\end{center}
\tablecomments{Results of our spectral index analysis for the single-pulses of our RRATs. $N_{8\sigma}$ is the number of single-pulses used in determining the mean spectral index, $\langle \alpha \rangle$, and thus were detected at a threshold of $8\sigma$. We report the mean spectral index with $1 \sigma$ errors.}
\end{table}

Additionally, a study by \cite{Karuppusamy2010} looked at the GRP spectral indices (both the main pulse and the interpulse) of the Crab pulsar. They fount that the distribution for both is also Gaussian and spans a range of spectral indices of about $-10$ to $+10$. This is much larger than the range of spectral indices we find for PSRs~J1819$-$1458, J1317$-$5759, and J1913$+$1330, but similar to that found for FRB~121102, which ranges from $-10$ to $+14$. The range of our RRAT spectral indices falls in between those found by \cite{Kramer2003} and \cite{Karuppusamy2010}, but with such a dearth of single-pulse spectral index studies, we cannot say whether the range we find is unusual. 

Many studies have looked at the distributions of pulsar spectral indices \citep[e.g.][]{Lorimer1995, Maron2000, Bates2013, Jankowski2017}. The mean value of pulsar spectral indices has varied slightly with each analysis. \cite{Lorimer1995} reported a mean spectral index of $-1.6$ in a study of 280 pulsars; \cite{Maron2000} reported a mean of $-1.8 \pm 0.2$ in a study of 281 pulsars, where the uncertainty is the error on the mean from individual spectral index measurements. \cite{Bates2013} reported a mean of $-1.4$ using Monte Carlo simulations, and \cite{Jankowski2017} reporting a weighted mean of $-1.60 \pm 0.03$ in a study of 441 pulsars where the uncertainty is the error on the weighted mean from individual spectral index measurements. 

While individually each RRAT falls within the standard spread of mean pulsar spectral indices, which can range from steeper than $-3$ to flatter than $-1$ \citep[e.g., see][]{Lorimer1995, Jankowski2017}, our mean RRAT spectral index is $-1.1 \pm 0.2$ and so is $2\sigma$ or more away from most previous values. While the methods of our spectral index calculations are different than those traditionally used \citep[e.g.][]{Lorimer1995,Jankowski2017}, all three RRATs exhibit flatter spectra at 1400~MHz than might be expected from these previous pulsar spectral index studies, but are similar to magnetar spectral indices \citep[e.g.][]{Camilo2008,Pennucci2015}. 

We also note that in no single pulse do we see the same narrow-band frequency structure that is seen in FRB~121102 \citep{Spitler2016,Gajjar2018,Michilli2018}. As the majority of pulses analyzed for all three RRATs tend toward having negative spectral indices distributed normally around the mean, standard pulsar emission mechanisms seem unlikely to be the source of FRB~121102. Neither the frequency structure of the pulses nor the spectral index values of the pulses exhibit the characteristics of FRB~121102.

Another interesting comparison is with giant radio pulses (GRPs) from the Crab pulsar. Many multi-frequency studies of the Crab pulsar have found that the spectral index is relatively steep and varies widely. As mentioned before, \cite{Karuppusamy2010} find the single-pulse spectral index spread of the main component of the Crab pulsar's GRPs to range from about $-10$ to $+10$. Other studies such as \cite{Popov2007} found that the spectral index of the main component of the Crab pulsar's GRPs ranges from $\alpha = -1.7$ to $-3.2$ depending on the width of the pulse. In a wide-band study of Crab GRPs, \cite{Mikami2016} found that the spectral index of GRPs ranges from $-1$ to $-4$, which is in line with the spread of our spectral index distribution, but generally steeper than the mean spectral indices of our RRATs. Similarly, \cite{Meyers2017} found that at low frequencies, between 120 and 165~MHz, the Crab GRPs show a spectral flattening with $\alpha = -0.7 \pm 1.4$, but between 732 and 3100~MHz $\alpha = -2.6 \pm 0.5$. This is steeper than the mean spectral index values for our three RRATs at 1400~MHz and suggests that the RRAT pulses, although bright and sporadic, have emission mechanisms similar to normal pulsars.

\subsection{Wait Times} \label{WTDisc}

\subsubsection{Wait-time Distribution Analysis} \label{WTDistAnalysis}

\begin{figure}[t]
\centering
\includegraphics[width=\columnwidth]{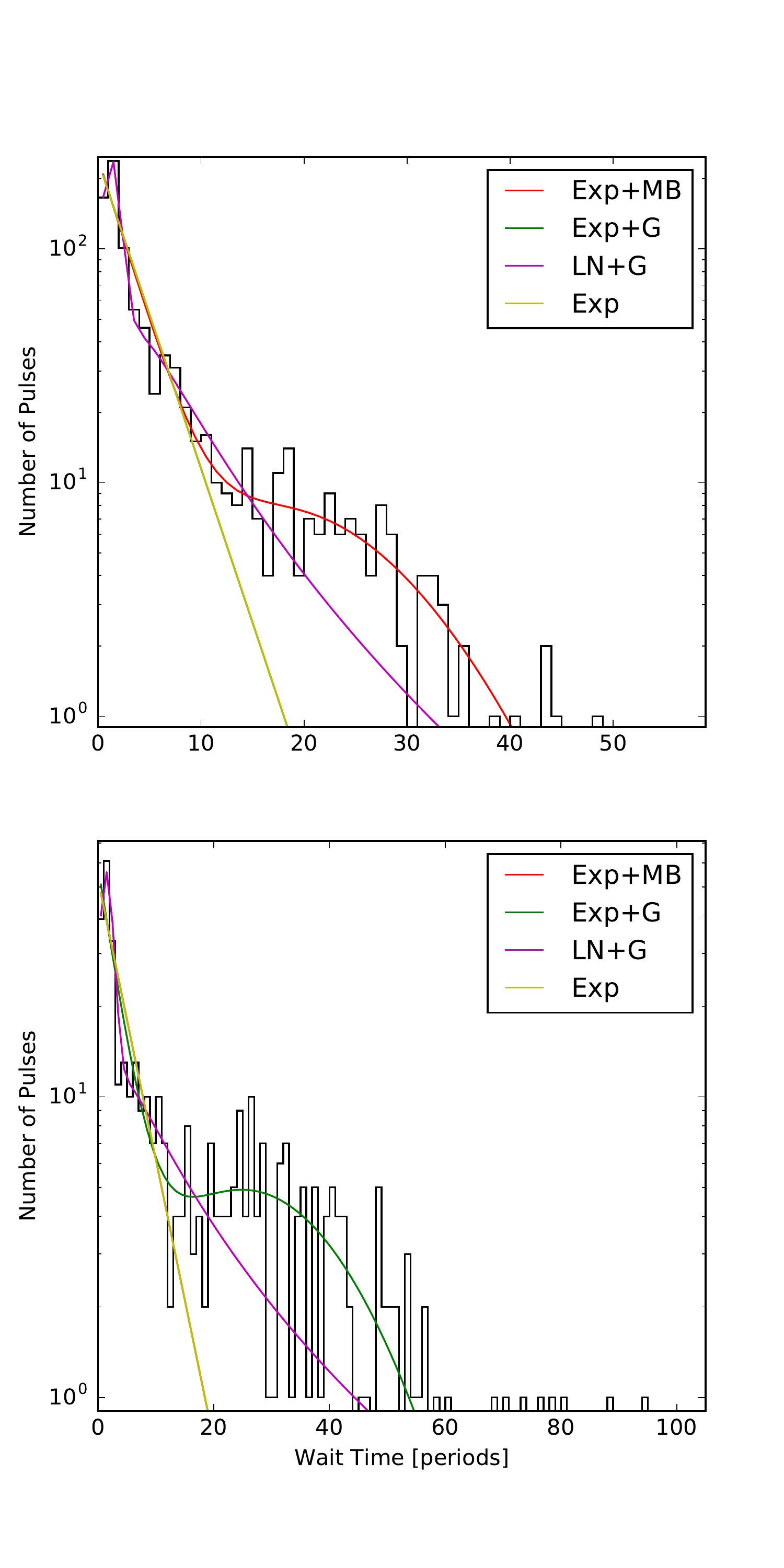}
\caption{Upper panel: distributions of single-pulse wait times for the GBT observations of PSR~J1819$-$1458 at S-band. We see that most pulses arrive within a few periods of each other. We also see that there is an extended tail of longer wait-times, but it does not go as far as that in Figure \ref{J1819ParkesWTHist}. The bump around 25 periods is less pronounced here than as seen in Figure \ref{J1819ParkesWTHist}. We fit the full distribution with an exponential plus a Gaussian (Exp+G; green), an exponential plus a Maxwell-Boltzmann distribution (Exp+MB; red), a log-normal plus a Gaussian (LN+G; magenta), and a pure exponential (Exp; yellow). Lower panel: same as the upper panel but we have used only pulses with an S/N of $>20\sigma$, approximately the equivalent detection threshold of Parkes at the L-band. We see that we clearly recover the bump around 25 periods and can detect longer wait-times. \label{J1819GBTWTHist}}
\end{figure}

\begin{figure*}[!p]
\centering
\includegraphics[width=\textwidth]{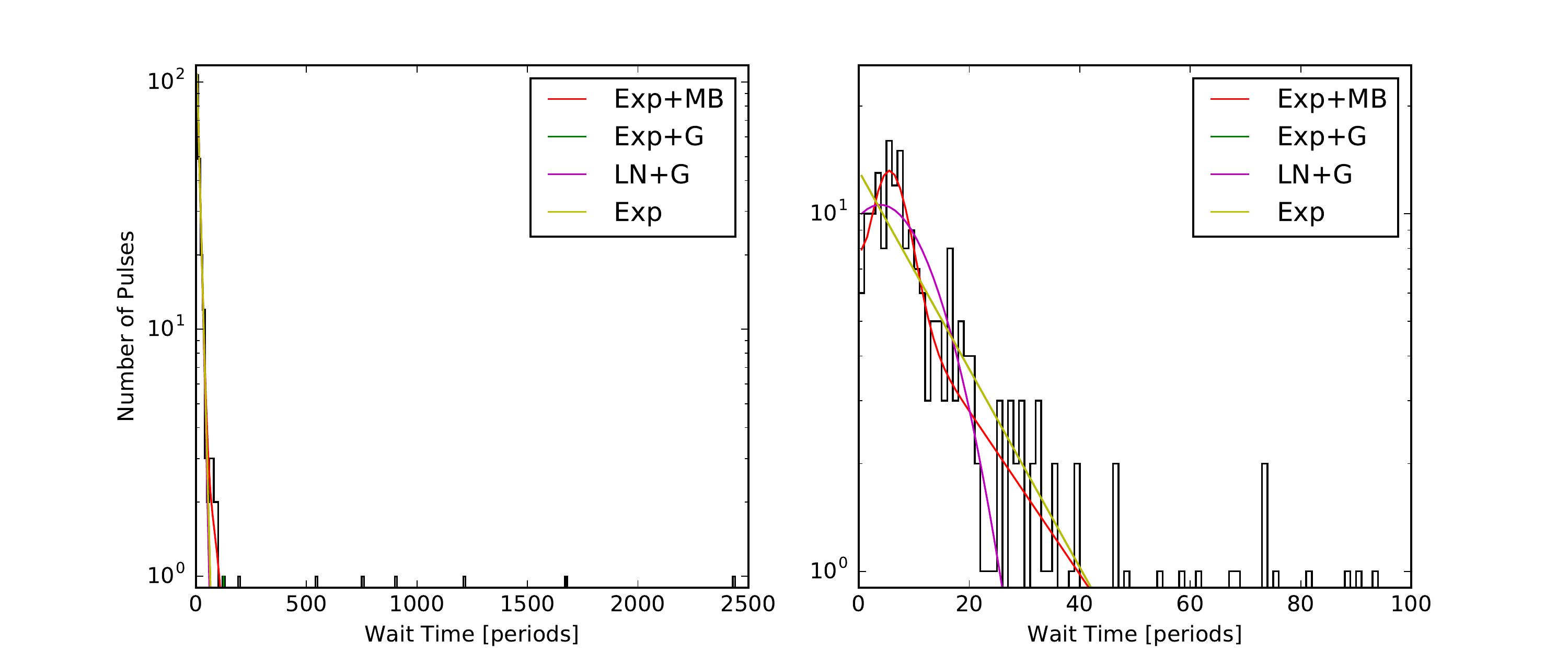}
\caption{Left panel: distributions of single-pulse wait times for PSR~J1913$+$1330 as observed with the Parkes Radio Telescope at the L-band. We have binned these pulses by 10 times their period and see that almost all pulses have shorter wait-times but there is an extended tail of long wait-times. We fit the full distribution with an exponential plus a Gaussian (Exp+G; green), an exponential plus a Maxwell-Boltzmann distribution (Exp+MB; red), a log-normal plus a Gaussian (LN+G; magenta), and a pure exponential (Exp; yellow). Right panel: same as the left panel but we have zoomed in on the first 100 periods and binned by the period. We see that most pulses appear within a few periods of each other. \label{J1913WTHist}}
\end{figure*}

\begin{figure*}[!p]
\centering
\includegraphics[width=\textwidth]{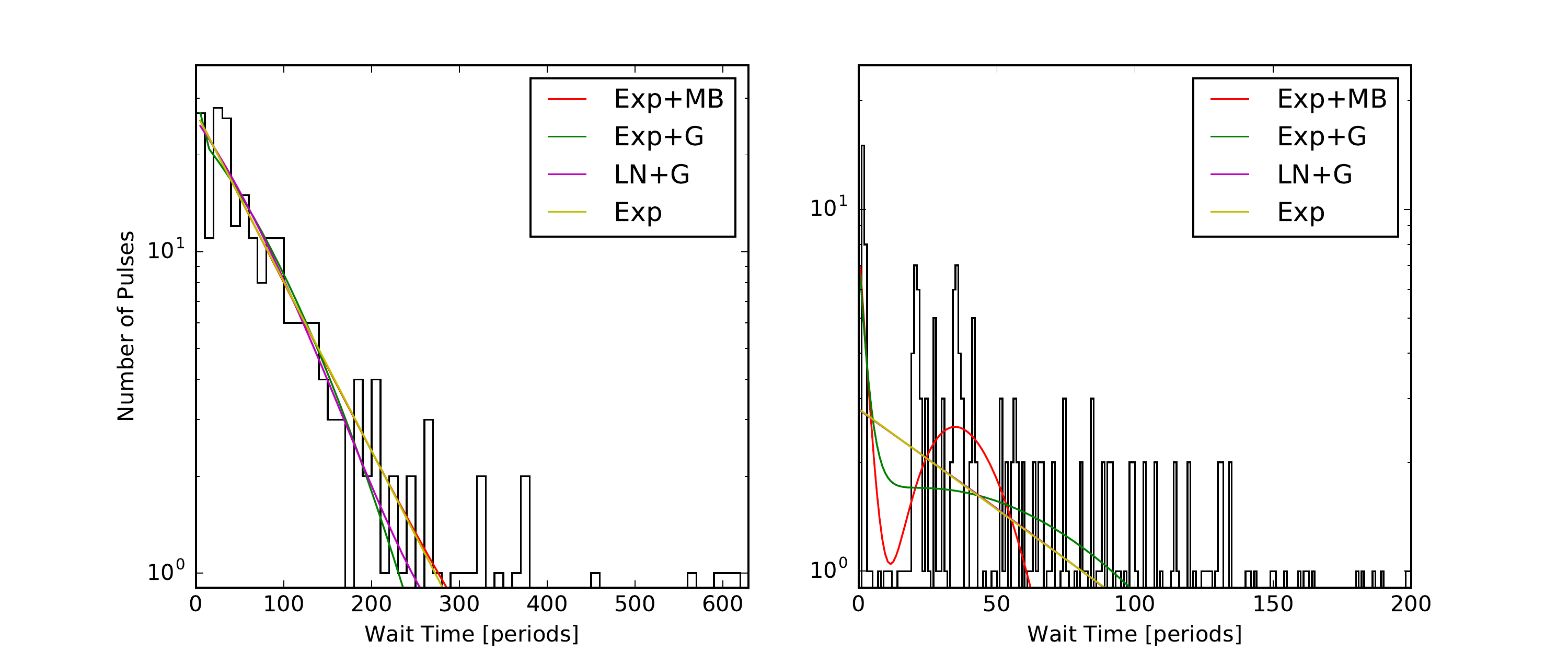}
\caption{Left panel: distributions of single-pulse wait times for PSR~J1317$-$5759 as observed by the Parkes Radio Telescope at the L-band. We have binned these pulses by 10 times their period and see that almost all pulses have shorter wait-times but there is a clear non-exponential tail of longer wait-times. We fit the full distribution with an exponential plus a Gaussian (Exp+G; green), an exponential plus a Maxwell-Boltzmann distribution (Exp+MB; red), a log-normal plus a Gaussian (LN+G; magenta), and a pure exponential (Exp; yellow). Right panel: same as the left panel but we have zoomed in on the first 200 periods and binned by the period. We see that most pulses appear within a few periods of each other but there is a second peak around 25 periods. \label{J1317WTHist}}
\end{figure*}

We first analyzed the wait-time distributions, seen in Figures \ref{J1819ParkesWTHist}, \ref{J1819GBTWTHist}, \ref{J1913WTHist}, and \ref{J1317WTHist}. If the emission was purely random, then the distribution would follow a pure exponential. We have attempted to model this dual population as a pure exponential, an exponential plus a Gaussian, an exponential plus a Maxwell-Boltzmann distribution, and as a log-normal plus a Gaussian. These processes are described in \S \ref{WTmethod}. In the bottom panels, we have fit only the portion of the distribution shown.

We find that for all three RRATs, most pulses are emitted within a few periods of each other, as expected for a random distribution. For PSR~J1317$-$5759 and the Parkes observations of PSR~J1819$-$1458, there is a secondary component around wait-times of $\sim 25$ periods ($\sim 106$~s for PSR~J1819$-$1458 and $\sim 66$~s for PSR~J1317$-$5759). The increase in number of pulses around 25~periods, and the extended tails of longer wait-times for PSR~J1317$-$5759 and the Parkes observations of PSR~J1819$-$1458, are clearly not exponential. While these extended tails are reminiscent of a Weibull distribution, similar to a Poisson distribution but with an extra shape parameter \citep{Oppermann2018}, this cannot describe both the peak of pulses within a few periods and the secondary component observed in the wait-time distributions, and we do not fit for it.

\afterpage{}

For PSR~J1913$+$1330, the wait-time distribution does not have the same increase in the number of pulses around wait-times of $\sim 25$~periods. However we note that PSR~J1913$+$1330 has both the fewest number of detected pulses and the longest wait-times, extending out to almost the full length of the observation in some cases. Furthermore, we note that the GBT observations of PSR~J1819$-$1458 do not show an extended tail of longer wait-times, and the increase in pulse number around 25~periods is much less pronounced. 

\begin{table*}
\begin{center}
\caption{Wait-time Model Statistics \label{WTStats}}
\begin{tabular}{c c c c c c c}
\tableline
RRAT & Model & DoF & $\chi^{2}$ & $\chi^{2}_{r}$ & F-Stat & F-Sig \\
\tableline
PSR~J1819$-$1458 (Parkes) & Exponential & 93 & 466 & 5.01 & - & - \\
             & Exponential and Maxwell-Boltzmann & 91 & 181 & 1.99 & 70.8 & $1.1\times 10^{-16}$ \\
             & Exponential and Gaussian & 90 & 182 & 2.03 & 46.1 & $1.1 \times 10^{-16}$ \\
             & Log Normal & 92 & 1070 & 11.63 & - & - \\
             & Log Normal and Gaussian & 89 & 334 & 3.76 & 64.4 & $1.1\times 10^{-16}$ \\
\tableline
PSR~J1819$-$1458 (GBT) & Exponential & 62 & 181 & 2.91 & - & - \\
             & Exponential and Maxwell-Boltzmann & 60 & 180 & 3.01 & - & - \\
             & Exponential and Gaussian & 59 & 120 & 2.03 & 9.80 & $2.4 \times 10^{-5}$ \\
             & Log Normal & 61 & 133 & 2.18 & - & - \\
             & Log Normal and Gaussian & 58 & 80 & 1.38 & 12.58 & $1.9 \times 10^{-6}$ \\
\tableline
PSR~J1317$-$5759 & Exponential & 128 & 112 & 0.87 & - & - \\
             & Exponential and Maxwell-Boltzmann & 126 & 144 & 1.14 & - & - \\
             & Exponential and Gaussian & 125 & 106 & 0.85 & 2.06 & $0.19$ \\
             & Log Normal & 127 & 121 & 0.95 & - & - \\
             & Log Normal and Gaussian & 124 & 111 & 0.90 & 3.61 & $1.5 \times 10^{-2}$ \\
\tableline
PSR~J1913$+$1330 & Exponential & 56 & 56 & 1.01 & - & - \\
             & Exponential and Maxwell-Boltzmann & 54 & 39 & 0.72 & 12.17 & $4.3 \times 10^{-5}$ \\
             & Exponential and Gaussian & 53 & 55 & 1.04 & 0.31 & $0.82$ \\
             & Log Normal & 55 & 44 & 0.79 & - & - \\
             & Log Normal and Gaussian & 52 & 62 & 1.20 & - & - \\
\tableline
\end{tabular}
\end{center}
\tablecomments{Model-fitting statistics for the wait-time distributions. DoF is degrees of freedom of the model, $\chi^{2}$ is the chi-squared value, and $\chi^{2}_{r}$ is the reduced chi-squared value. The F-statistic and significance correspond to the comparison of a more complex model with the less complex corresponding model. If there is no F-statistic reported, either the model is one of the base models, or the chi-squared value of the more complex model is larger than the base model. }
\end{table*}

A partial explanation of this secondary component can be found by considering the sensitivity of the GBT at the S-band to Parkes at the L-band. The GBT at 2~GHz should be around four times more sensitive than Parkes at 1400~MHz. We can make a direct comparison between the two distributions if we only include single-pulses from the GBT observation that are bright enough have been detected by Parkes at 1400~MHz. This filtered distribution is shown in the bottom panel of Figure \ref{J1819GBTWTHist} and shows the extended tail of wait-times, out to 100 periods, and recovers the slight bulge in number of pulses around 25 periods. Despite the lower number, this modulation of bright pulses around $25$~periods is evidence that RRATs may exhibit short-timescale emission trends, contrary to \cite{Palliyaguru2011}. Since we count wait-times between the pulses in a single observation, we do not see the long-timescale trends observed by \cite{Palliyaguru2011} as the the shortest trend they see in our three RRATs is for PSR~J1317$-$5759 at 1.9 hr, longer than almost all of our individual observations.

Even accounting for this bias, for the full PSR~J1819$-$1458 wait-time distribution, we see two populations of emission, one accounting for the pulses in each burst, described by the steeper part of the distribution with wait-times of a few periods, and another that accounts for the distribution of the bursts, described by a secondary distribution component ``bump'' around 25 periods. These distribution can also be seen in the wait-time distributions of PSR~J1317$-$5759, shown in Figure \ref{J1317WTHist}. 

In addition to the $\chi^{2}_{r}$, we have also compared distributions using an F-test where appropriate. The F-statistic for model regression tells us if adding more parameters to our model makes a statistically significant contribution to the fit. The F-test assumes that the more complicated model has a smaller $\chi^{2}$ value, and is thus computed by
\begin{equation} \label{Ftest}
F = \frac{(\chi^{2}_{1}-\chi^{2}_{2})/(p_{2}-p_{1})}{(\chi^{2}_{2})/(n-p_{2}-1)} .
\end{equation}
Here $\chi^{2}_{1/2}$ and $p_{1/2}$ are the chi-squared and number of free parameters that describe each model, where model~2 is the more complex model, and $n$ is the number of data points used to fit the models. If the $\chi^{2}$ of model~2 is greater than that of model~1, we do not compute the F-statistic as this means the less complex model better fits the data. 

\afterpage{}
\begin{figure*}
\includegraphics[width=\textwidth]{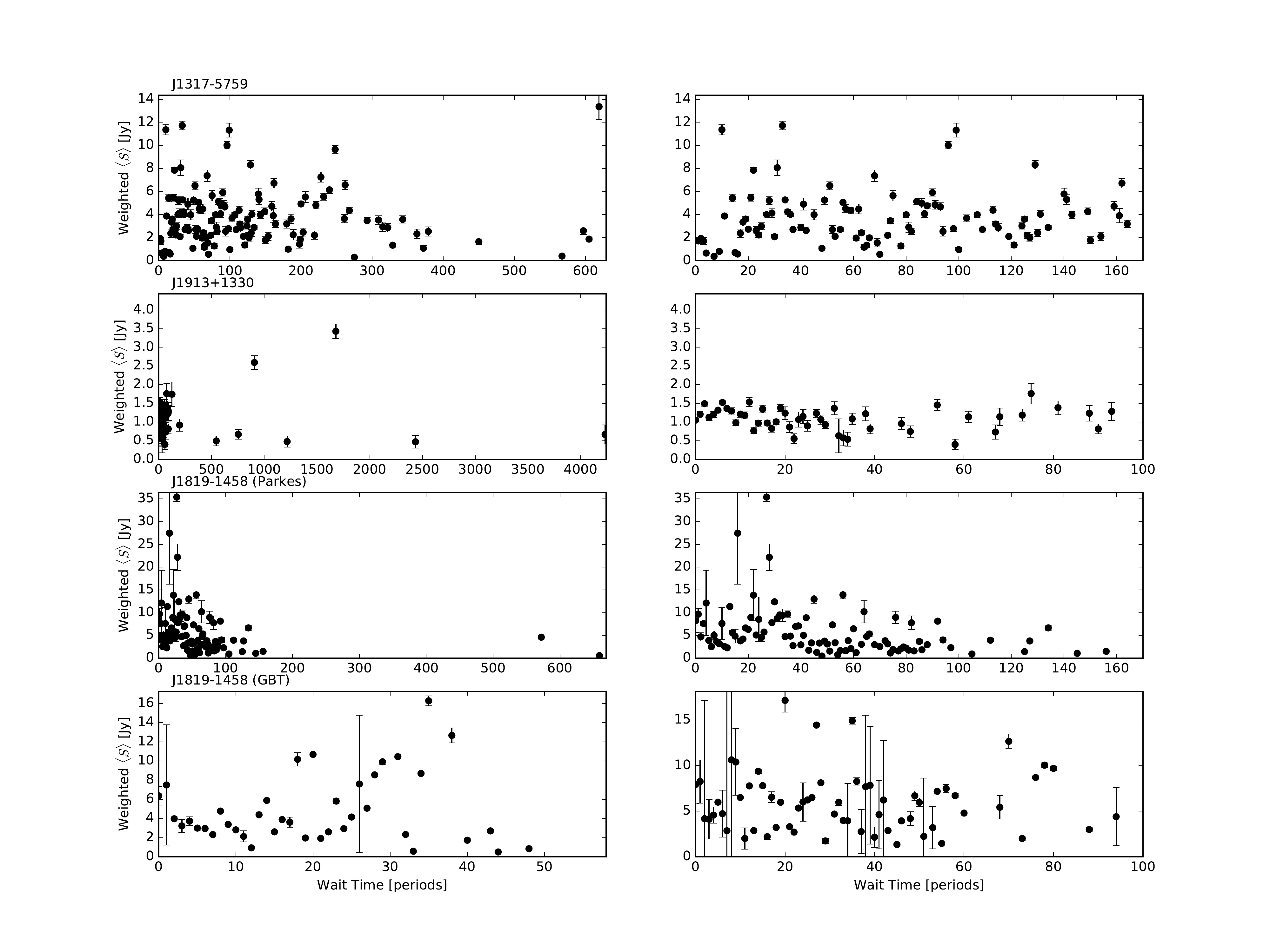}
\caption{Weighted mean flux density of the single-pulses from each RRAT vs. wait time between pulses. The top three right panels are insets of the top three left panels. The right-hand bottom panel is not a inset but has the pulses that are below the approximate detection threshold of Parkes at the L-band filtered out. The RRAT corresponding to each set of panels is denoted in the upper left of the left-hand panels. We see that for all RRATs, the wait time does not correlate with flux density. \label{AllWTvI}}
\end{figure*}

\begin{figure*}
\centering
\includegraphics[width=\textwidth]{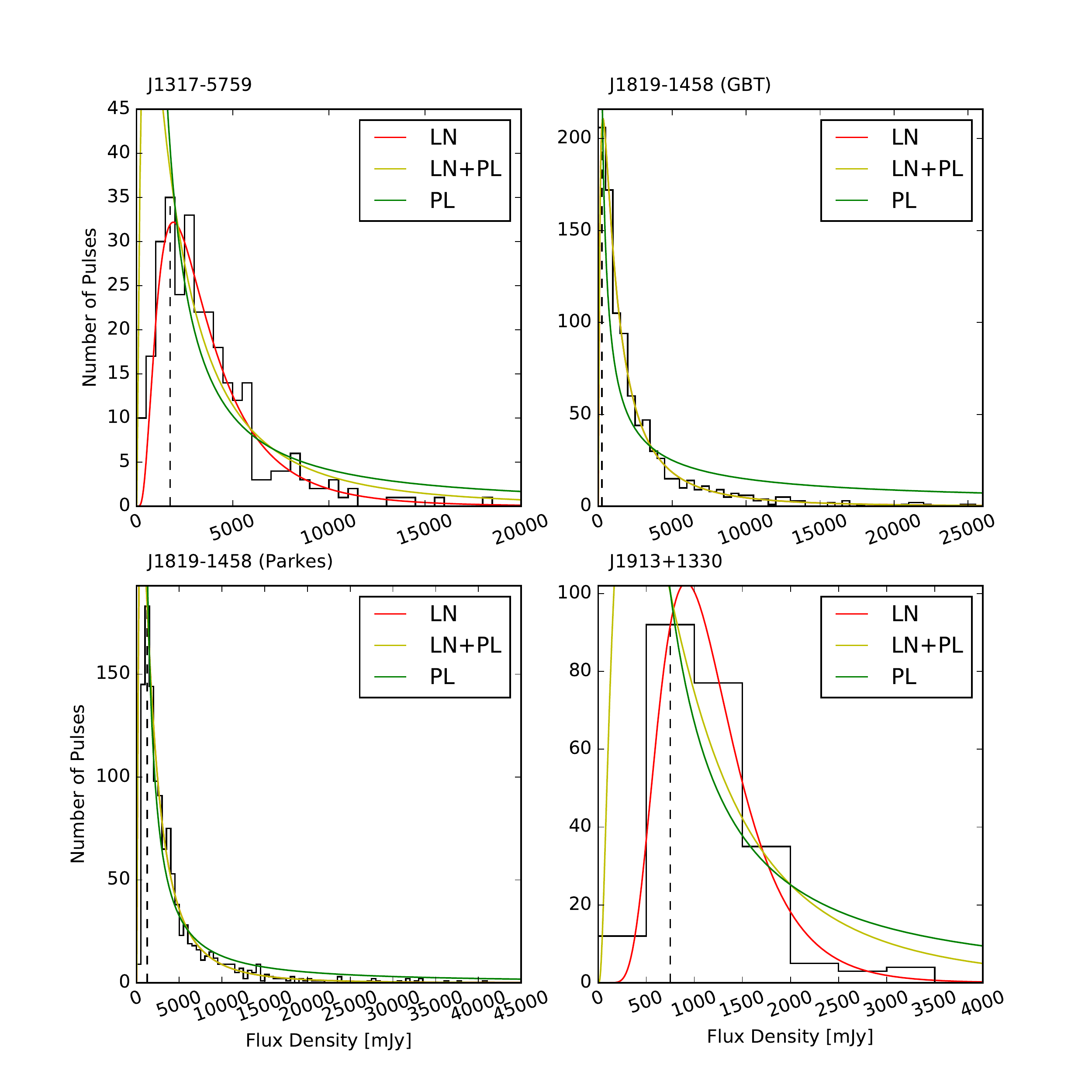}
\caption{Distribution of single-pulse total flux density for each RRAT in our analysis. The RRAT corresponding to each distribution is denoted in above each plot at the upper left. Each distribution is fit with a log-normal (LN; red), power law (PL; green) or combination (LN+PL; yellow). We fit the distribution only to the dashed black line to reduce the influence of the non-detection of weak pulses. \label{AmpDistsPlot}}
\end{figure*}

We can determine the significance of the model from the Cumulative Density Function obtained using the \verb+scipy+ function \verb+fdtr+. These values, along with the degrees of freedom (DoF), $\chi^{2}$, and $\chi^{2}_{r}$ of each model, are reported in Table \ref{WTStats}. We report these values only for the ``zoomed-in'' section of each wait-time distribution. Our models are unable to fit the extended wait-time tails of the full distributions well, showing that there could be some other, more complex emission process not well fit by our simple models.

From the statistics reported in Table \ref{WTStats}, it is clear that none of our models has produced a good fit for PSR~J1819$-$1458 or PSR~J1317$-$5759. The Parkes observations of PSR~J1819$-$1458 show that adding extra parameters to any of our models makes a significant improvement to our fits, but does not tell us if one is better than another. The GBT observations of PSR~J1819$-$1458 show that adding a Gaussian to either a power law or log-normal is statistically significant, but again we cannot tell which model is preferred. The F-statistics for PSR~J1317$-$5759 favor the two simplest models, and we cannot determine if an exponential better fits the wait-time distribution than a log-normal. This likely means that a more complex model is needed to fit these wait-time distributions. For PSR~J1913$+$1330, we find that both a power law plus a Gaussian and a pure power law fit the distribution well, but adding the Gaussian is not statistically significant.

While the wait-time distribution of FRB~121102 is well described by a Weibull distribution \citep{Oppermann2018}, it is not as apt for the RRAT wait-time distributions. The latter shows the secondary peaks which are not seen in FRB~121102 and cannot be accounted for by the Weibull distribution although it can account for the extended tails of the long wait-times seen for the RRATs.

For PSR~J1913$+$1330, an exponential distribution seems to best describe the wait-time distribution, which supports a random emission model. As we do not see a secondary component in the wait-time distribution of PSR~J1913$+$1330, this seems reasonable. However it is likely that, as with PSR~J1819$-$1458 and PSR~J1317$-$5759, we are still missing some pulses. Weak pulse analyses have been performed for RRATs \citep[e.g.][]{Cui2017, Jiang2017} and recent work by \cite{Bhattacharyya2018} has found that PSR~J1913$+$1330 exhibits a weak mode emission followed by long periods where there is no detectable emission. A lack of bright pulses could explain why our pulse wait-time distribution appears to support purely random emission, and does not show the secondary component seen in PSR~J1819$-$1458 and PSR~J1317$-$5759.

\begin{table*}
\begin{center}
\caption{Energy Model Statistics \label{AmplitudeStats}}
\begin{tabular}{c c c c c c c}
\tableline
RRAT & Model & DoF & $\chi^{2}$ & $\chi^{2}_{r}$ & F-Stat & F-Sig \\
\tableline
PSR~J1819$-$1458 (Parkes) & Log normal & 51 & 43.3 & 0.85 & \multirow{2}{*}{$7.66 \times 10^{-3}$}  & \multirow{2}{*}{0.99} \\
             & Log normal and power law & 49 & 43.2 & 0.88 & \multirow{2}{*}{94.2}  & \multirow{2}{*}{$1.11 \times 10^{-16}$} \\
             & Power law  & 52 & 298 & 5.74 & & \\
\tableline
PSR~J1819$-$1458 (GBT) & Log normal & 34 & 22.4 & 0.66 & \multirow{2}{*}{$1.69 \times 10^{-5}$}  & \multirow{2}{*}{0.99} \\
             & Log normal and power law & 32 & 22.5 & 0.70 & \multirow{2}{*}{380}  & \multirow{2}{*}{$1.11 \times 10^{-16}$} \\
             & Power law  & 35 & 849 & 24.3 & & \\
\tableline
PSR~J1317$-$5759 & Log normal & 22 & 20.5 & 0.93 & \multirow{2}{*}{-}  & \multirow{2}{*}{-} \\
             & Log normal and power law & 20 & 33.96 & 1.70 & \multirow{2}{*}{4.07}  & \multirow{2}{*}{$2.06\times 10^{-2}$} \\
             & Power law & 23 & 55.8 & 2.43 & & \\
\tableline
PSR~J1913$+$1330 & Log normal & 3 & 83.9 & 2.80 & \multirow{2}{*}{-}  & \multirow{2}{*}{-} \\
             & Log normal and power law & 1 & 87.9 & 87.9 & \multirow{2}{*}{-}  & \multirow{2}{*}{-} \\
             & Power law  & 4 & 141 & 35.2 & & \\
\tableline
\end{tabular}
\end{center}
\tablecomments{Model-fitting statistics for the RRAT energy distributions. DoF is degrees of freedom of the model, $\chi^{2}$ is the chi-squared value, and $\chi^{2}_{r}$ is the reduced chi-squared value. The top F-statistic is a comparison of the power law and the combined power law and log-normal fit. The bottom value is for the log-normal and the combined power law and log-normal fit. F-significance is the significance of adding the fit parameters for a more complex fit.}
\end{table*}

\subsubsection{Flux-density-Wait-time Correlation} \label{FluxWTCor}

In addition we analyzed the total integrated flux density of each pulse as a function of wait-time for each RRAT, including both PSR~J1819$-$1458 observations, which can be seen in Figure \ref{AllWTvI}. For each RRAT we have binned the pulses by wait-time in units of the RATT period and computed the weighted mean of the total integrated pulse flux density per wait-time. We find that for the three RRATs analyzed there is no correlation between the wait-time and the flux density of the pulse. We therefore conclude that the emission is not a consequence of a process through which energy is ``stored up,'' such as magnetic recombination in the pulsar magnetosphere \citep[e.g.][]{Lyutikov2002, Rutledge2006}.

\subsection{Energy Distributions} \label{AmpDists}

Finally we present the pulse energy distributions for PSR~J1819$-$1458, PSR~J1317$-$5759, and PSR~J1913$+$1330, shown in Figure \ref{AmpDistsPlot}. The different observations of PSR~J1819$-$1458 are denoted above their respective panels. We have included every pulse above a $5\sigma$ detection limit, filtered as with our wait-time analysis above. We note that, while we examine the pulse energy distributions, we have also compared these results to the pulse amplitude distributions and find they are similar.

\cite{Cui2017} have shown that the RRAT pulse amplitude distribution can be described by either a log-normal, as with pulsars, or a log-normal plus a power-law component, as in the power-law distribution of GRPs \citep[e.g.][]{Mickaliger2012}. However PSR~J1819$-$1458, PSR~J1913$+$1330, and PSR~J1317$-$5759 were not included in that analysis. Previously, \cite{McLaughlin2006} found the pulse amplitude distributions for PSR~J1819$-$1458 and PSR~J1317$-$5759 to be flatter, with a power-law index of $\sim 1$, than those of GRPs that show a power-law index of $\sim 2-3$. This difference in power-law index shows that the RRAT emission mechanism is likely separate from the emission mechanism of GRPs.

All three RRATs analyzed appear to show a population of pulses in the Parkes observations that have larger than standard flux densities and could be characteristic of a GRP power-law distribution \citep[e.g.][]{Mickaliger2012}. To determine whether these are true power-law tails, we have fit a log-normal, power law, and a combined log-normal and power law, to each distribution, shown in Figure \ref{AmpDistsPlot}. We also compute the F-statistic for these distributions where appropriate as described in \S \ref{WTDisc}. We report these values as well as the DoF, $\chi^{2}$, and $\chi^{2}_{r}$ values for these models for each RRAT in Table \ref{AmplitudeStats}.

For both observations of PSR~J1819$-$1458, we find that a log-normal best describes the pulse energy distribution, and that adding a power-law component does not improve the fit significantly. For both PSRs~J1317$-$5759 and J1913$+$1330 the pulse energy distribution is best described by a log-normal distribution. 

Recent work by \cite{Mickaliger2018} found that the pulse energy distribution for RRATs showed both low- and high-energy peaks, indicative of the RRAT bursting, that could be fit with two log-normal distributions. We do not see evidence for this in our pulse energy analysis; however our analysis spans many epochs whereas \cite{Mickaliger2018} analyzed only data taken by the PMPS survey, so it is possible that we have bridged this gap with a larger distribution of pulses.  

While it is possible that the emission mechanism for RRATs is similar to that of GRPs, we lack the detections to fit the expected power-law tail. However as the spectral index distribution discussed in \S \ref{SpecIdxDisc} also does not follow the expected values for GRPs, we conclude that RRAT emission is not consistent with GRP emission.

We see from the significance of the F-test in Table \ref{AmplitudeStats} that the pulse energy distributions for PSR~J1819$-$1458 in both the Parkes and GBT data are best described by a log-normal distribution. As expected with a mean spectral index of $-1.1 \pm 0.2$, where the uncertainty is from the mean of individual RRAT weighted means, the flux densities for the GBT pulses are weaker than those of the Parkes 1400~MHz distribution. However the sensitivity of the GBT at 2~GHz is greater than that of Parkes at 1400~MHz, which explains the larger number of pulses detected, and is evidence that the pulses we detect with Parkes are part of a broader distribution of pulses. 

\section{Conclusions} \label{Conclusion}

In this work we completed a single-pulse analysis of three RRATs, PSRs~J1819$-$1458, J1317$-$5759, and J1913$+$1330, based on over 11 years of timing observations from the Parkes Radio Telescope, as well as archival data from the PMPS and an additional 7.5 hr observation of PSR~J1819$-$1458 from the GBT. We have developed a method for determining the spectral indices of bright single-pulses and have shown them to be normally distributed around a mean spectral index that is comparable to the spectral indices of most pulsars. While there are few other single-pulse spectral index distributions, we find that our distribution is wider than those found by \cite{Kramer2003}, and narrower than those found by \cite{Karuppusamy2010}, although all are normally distributed.

We have shown that the pulsed emission is not uniformly distributed on small timescales, and even exhibits clustering around $\sim 25$ pulse periods in PSRs~J1819$-$1458 and J1317$-$5759. Additionally, for PSR~J1913$+$1330 we note that the extended wait-time tail is real; however we are unable to explain it. This is likely due to the intrinsic nature of the RRAT, similar to the processes in nulling pulsars \citep[e.g.][]{Wang2007}. External mechanisms, such as an asteroid belt around the RRAT \citep{Cordes2008}, cannot be responsible for modulation on such short timescales, but can have an effect on longer timescales as discussed in \cite{Palliyaguru2011}. For these three RRATs we have additionally shown that the time between the pulses is not correlated with the flux density of the pulse, thus the emission is not due to ``storing up" energy.

Additionally we have analyzed the pulse energy distribution of our three RRATs. We found that PSRs~J1317$-$5759 and J1913$+$1330 agree well with previous pulse energy and amplitude distribution studies done on other RRATs from \cite{Cui2017}, showing a log-normal pulsar-like distribution. For PSR~J1819$-$1458, we see that adding a power law to the log-normal distribution is not statistically significant. We have also found that these three RRATs no not exhibit a power-law tail in the pulse energy distributions, indicative of GRPs.

Neither the single-pulse narrow-band emission seen in FRB~121102, nor the narrowband frequency structure seen in FRB~170827 \citep{Farah2018} are seen in the RRAT pulses, which suggests that the emission is different from RRAT emission. Instead our RRAT pulses seem to be broadband, more like non-repeating FRBs\cprotect\footnote{see \verb+http://frbcat.org/+ for a full list \citep{Petroff2016}.}. Although the spread of RRAT single-pulse spectral indices seems to be similar to the spread of spectral indices from FRB~121102, without more pulsar or RRAT single-pulse spectral index analyses, it is difficult to determine if this spread hints at a common emission process or not.

Unfortunately, the number of RRATs used in this analysis is small due to the fact that many bright pulses are necessary for a robust statistical analysis. Further observations of these RRATs at other frequencies would allow us to see if they exhibit spectra that change with frequency as has been found with the Crab pulsar. Observations with more sensitive telescopes would allow us to detect enough single-pulses to perform a single-pulse spectral analysis for other RRATs. Further observations of other bright RRATs would also allow us to perform our analysis on a larger sample to further explore their spectral distribution.

\section*{Acknowledgements} \label{acknowledgements}

This work was supported by NSF Award OIA-1458952. M.A.M. and B.J.S. are members of the NANOGrav Physics Frontiers Center which is supported by NSF award 1430284. B.J.S. acknowledges support from West Virginia
University through the STEM Mountains of Excellence Fellowship. The Parkes radio telescope is part of the Australia Telescope National Facility which is funded by the Australian Government for operation as a National Facility managed by CSIRO. The Green Bank Observatory is a facility of the National Science Foundation operated under cooperative agreement by Associated Universities, Inc. 

\section*{Software} \label{software}

{\textit{Software}}: GPy (https://sheffieldml.github.io/GPy/), Scipy \citep{Jones2001}, SIGPROC \citep{Lorimer2011}


\bibliography{single_pulse}{}

\begin{thebibliography}{}
\expandafter\ifx\csname natexlab\endcsname\relax\def\natexlab#1{#1}\fi

\bibitem[{{Bates} {et~al.}(2013){Bates}, {Lorimer}, \& {Verbiest}}]{Bates2013}
{Bates}, S.~D., {Lorimer}, D.~R., \& {Verbiest}, J.~P.~W. 2013, \mnras, 431,
  1352

\bibitem[{{Bhandari} {et~al.}(2018){Bhandari}, {Keane}, {Barr}, {Jameson},
  {Petroff}, {Johnston}, {Bailes}, {Bhat}, {Burgay}, {Burke-Spolaor}, {Caleb},
  {Eatough}, {Flynn}, {Green}, {Jankowski}, {Kramer}, {Krishnan}, {Morello},
  {Possenti}, {Stappers}, {Tiburzi}, {van Straten}, {Andreoni}, {Butterley},
  {Chandra}, {Cooke}, {Corongiu}, {Coward}, {Dhillon}, {Dodson}, {Hardy},
  {Howell}, {Jaroenjittichai}, {Klotz}, {Littlefair}, {Marsh}, {Mickaliger},
  {Muxlow}, {Perrodin}, {Pritchard}, {Sawangwit}, {Terai}, {Tominaga}, {Torne},
  {Totani}, {Trois}, {Turpin}, {Niino}, {Wilson}, {Albert}, {Andr{\'e}},
  {Anghinolfi}, {Anton}, {Ardid}, {Aubert}, {Avgitas}, {Baret},
  {Barrios-Mart{\'{\i}}}, {Basa}, {Belhorma}, {Bertin}, {Biagi}, {Bormuth},
  {Bourret}, {Bouwhuis}, {Br{\^a}nza{\c s}}, {Bruijn}, {Brunner}, {Busto},
  {Capone}, {Caramete}, {Carr}, {Celli}, {Moursli}, {Chiarusi}, {Circella},
  {Coelho}, {Coleiro}, {Coniglione}, {Costantini}, {Coyle}, {Creusot},
  {D{\'{\i}}az}, {Deschamps}, {De Bonis}, {Distefano}, {Palma}, {Domi},
  {Donzaud}, {Dornic}, {Drouhin}, {Eberl}, {Bojaddaini}, {Khayati},
  {Els{\"a}sser}, {Enzenh{\"o}fer}, {Ettahiri}, {Fassi}, {Felis}, {Fusco},
  {Gay}, {Giordano}, {Glotin}, {Gregoire}, {Gracia-Ruiz}, {Graf}, {Hallmann},
  {van Haren}, {Heijboer}, {Hello}, {Hern{\'a}ndez-Rey}, {H{\"o}{\ss}l},
  {Hofest{\"a}dt}, {Hugon}, {Illuminati}, {James}, {de Jong}, {Jongen},
  {Kadler}, {Kalekin}, {Katz}, {Kie{\ss}ling}, {Kouchner}, {Kreter},
  {Kreykenbohm}, {Kulikovskiy}, {Lachaud}, {Lahmann}, {Lef{\`e}vre}, {Leonora},
  {Loucatos}, {Marcelin}, {Margiotta}, {Marinelli}, {Mart{\'{\i}}nez-Mora},
  {Mele}, {Melis}, {Michael}, {Migliozzi}, {Moussa}, {Navas}, {Nezri},
  {Organokov}, {P{\v a}v{\v a}la{\c s}}, {Pellegrino}, {Perrina}, {Piattelli},
  {Popa}, {Pradier}, {Quinn}, {Racca}, {Riccobene}, {S{\'a}nchez-Losa},
  {Salda{\~n}a}, {Salvadori}, {Samtleben}, {Sanguineti}, {Sapienza},
  {Sch{\"u}ssler}, {Sieger}, {Spurio}, {Stolarczyk}, {Taiuti}, {Tayalati},
  {Trovato}, {Turpin}, {T{\"o}nnis}, {Vallage}, {Van Elewyck}, {Versari},
  {Vivolo}, {Vizzocca}, {Wilms}, {Zornoza}, \& {Z{\'u}{\~n}iga}}]{Bhandari2018}
{Bhandari}, S., {Keane}, E.~F., {Barr}, E.~D., {et~al.} 2018, \mnras, 475, 1427

\bibitem[{{Bhat} {et~al.}(2004){Bhat}, {Cordes}, {Camilo}, {Nice}, \&
  {Lorimer}}]{Bhat2004}
{Bhat}, N.~D.~R., {Cordes}, J.~M., {Camilo}, F., {Nice}, D.~J., \& {Lorimer},
  D.~R. 2004, \apj, 605, 759

\bibitem[{{Bhattacharyya} {et~al.}(2018){Bhattacharyya}, {Lyne}, {Stappers},
  {Weltevrede}, {Keane}, {McLaughlin}, {Kramer}, {Jordan}, \&
  {Bassa}}]{Bhattacharyya2018}
{Bhattacharyya}, B., {Lyne}, A.~G., {Stappers}, B.~W., {et~al.} 2018, \mnras,
  477, 4090

\bibitem[{{Burke-Spolaor} {et~al.}(2012){Burke-Spolaor}, {Johnston}, {Bailes},
  {Bates}, {Bhat}, {Burgay}, {Champion}, {D'Amico}, {Keith}, {Kramer}, {Levin},
  {Milia}, {Possenti}, {Stappers}, \& {van Straten}}]{BS2012}
{Burke-Spolaor}, S., {Johnston}, S., {Bailes}, M., {et~al.} 2012, \mnras, 423,
  1351

\bibitem[{{Camilo} {et~al.}(2008){Camilo}, {Reynolds}, {Johnston}, {Halpern},
  \& {Ransom}}]{Camilo2008}
{Camilo}, F., {Reynolds}, J., {Johnston}, S., {Halpern}, J.~P., \& {Ransom},
  S.~M. 2008, \apj, 679, 681

\bibitem[{{Chatterjee} {et~al.}(2017){Chatterjee}, {Law}, {Wharton},
  {Burke-Spolaor}, {Hessels}, {Bower}, {Cordes}, {Tendulkar}, {Bassa},
  {Demorest}, {Butler}, {Seymour}, {Scholz}, {Abruzzo}, {Bogdanov}, {Kaspi},
  {Keimpema}, {Lazio}, {Marcote}, {McLaughlin}, {Paragi}, {Ransom}, {Rupen},
  {Spitler}, \& {van Langevelde}}]{Chatterjee2017}
{Chatterjee}, S., {Law}, C.~J., {Wharton}, R.~S., {et~al.} 2017, \nat, 541, 58

\bibitem[{{Cordes} {et~al.}(2004){Cordes}, {Bhat}, {Hankins}, {McLaughlin}, \&
  {Kern}}]{Cordes2004}
{Cordes}, J.~M., {Bhat}, N.~D.~R., {Hankins}, T.~H., {McLaughlin}, M.~A., \&
  {Kern}, J. 2004, \apj, 612, 375

\bibitem[{{Cordes} \& {Lazio}(2002)}]{NE2001}
{Cordes}, J.~M., \& {Lazio}, T.~J.~W. 2002, ArXiv Astrophysics e-prints,
  astro-ph/0207156

\bibitem[{{Cordes} \& {Rickett}(1998)}]{Cordes1998}
{Cordes}, J.~M., \& {Rickett}, B.~J. 1998, \apj, 507, 846

\bibitem[{{Cordes} \& {Shannon}(2008)}]{Cordes2008}
{Cordes}, J.~M., \& {Shannon}, R.~M. 2008, \apj, 682, 1152

\bibitem[{{Cui} {et~al.}(2017){Cui}, {Boyles}, {McLaughlin}, \&
  {Palliyaguru}}]{Cui2017}
{Cui}, B.-Y., {Boyles}, J., {McLaughlin}, M.~A., \& {Palliyaguru}, N. 2017,
  \apj, 840, 5

\bibitem[{{Farah} {et~al.}(2018){Farah}, {Flynn}, {Bailes}, {Jameson},
  {Bannister}, {Barr}, {Bateman}, {Bhandari}, {Caleb}, {Campbell-Wilson},
  {Chang}, {Deller}, {Green}, {Hunstead}, {Jankowski}, {Keane}, {Macquart},
  {M{\"o}ller}, {Onken}, {Os{\l}owski}, {Parthasarathy}, {Plant}, {Ravi},
  {Shannon}, {Tucker}, {Venkatraman Krishnan}, \& {Wolf}}]{Farah2018}
{Farah}, W., {Flynn}, C., {Bailes}, M., {et~al.} 2018, \mnras, 478, 1209

\bibitem[{{Gajjar} {et~al.}(2018){Gajjar}, {Siemion}, {Price}, {Law},
  {Michilli}, {Hessels}, {Chatterjee}, {Archibald}, {Bower}, {Brinkman},
  {Burke-Spolaor}, {Cordes}, {Croft}, {Enriquez}, {Foster}, {Gizani},
  {Hellbourg}, {Isaacson}, {Kaspi}, {Lazio}, {Lebofsky}, {Lynch}, {MacMahon},
  {McLaughlin}, {Ransom}, {Scholz}, {Seymour}, {Spitler}, {Tendulkar},
  {Werthimer}, \& {Zhang}}]{Gajjar2018}
{Gajjar}, V., {Siemion}, A.~P.~V., {Price}, D.~C., {et~al.} 2018, ArXiv
  e-prints, arXiv:1804.04101

\bibitem[{{Haslam} {et~al.}(1981){Haslam}, {Klein}, {Salter}, {Stoffel},
  {Wilson}, {Cleary}, {Cooke}, \& {Thomasson}}]{Haslam1981}
{Haslam}, C.~G.~T., {Klein}, U., {Salter}, C.~J., {et~al.} 1981, \aap, 100, 209

\bibitem[{{Jankowski} {et~al.}(2017){Jankowski}, {van Straten}, {Keane},
  {Bailes}, {Barr}, {Johnston}, \& {Kerr}}]{Jankowski2017}
{Jankowski}, F., {van Straten}, W., {Keane}, E.~F., {et~al.} 2017, ArXiv
  e-prints, arXiv:1709.08864

\bibitem[{{Jiang} {et~al.}(2017){Jiang}, {Cui}, {Schmid}, {McLaughlin}, \&
  {Cao}}]{Jiang2017}
{Jiang}, M., {Cui}, B.-Y., {Schmid}, N.~A., {McLaughlin}, M.~A., \& {Cao},
  Z.-C. 2017, \apj, 847, 75

\bibitem[{Jones {et~al.}(2001)Jones, Oliphant, Peterson, {et~al.}}]{Jones2001}
Jones, E., Oliphant, T., Peterson, P., {et~al.} 2001, {SciPy}: Open source
  scientific tools for {Python}, [Online; accessed <today>]

\bibitem[{{Karastergiou} {et~al.}(2009){Karastergiou}, {Hotan}, {van Straten},
  {McLaughlin}, \& {Ord}}]{Karastergiou2009}
{Karastergiou}, A., {Hotan}, A.~W., {van Straten}, W., {McLaughlin}, M.~A., \&
  {Ord}, S.~M. 2009, \mnras, 396, L95

\bibitem[{{Karuppusamy} {et~al.}(2010){Karuppusamy}, {Stappers}, \& {van
  Straten}}]{Karuppusamy2010}
{Karuppusamy}, R., {Stappers}, B.~W., \& {van Straten}, W. 2010, \aap, 515, A36

\bibitem[{{Keane}(2016)}]{Keane2016}
{Keane}, E.~F. 2016, \mnras, 459, 1360

\bibitem[{{Keane} {et~al.}(2011){Keane}, {Kramer}, {Lyne}, {Stappers}, \&
  {McLaughlin}}]{Keane2011}
{Keane}, E.~F., {Kramer}, M., {Lyne}, A.~G., {Stappers}, B.~W., \&
  {McLaughlin}, M.~A. 2011, \mnras, 415, 3065

\bibitem[{{Keane} \& {Petroff}(2015)}]{Keane2015}
{Keane}, E.~F., \& {Petroff}, E. 2015, \mnras, 447, 2852

\bibitem[{{Kramer} {et~al.}(2003){Kramer}, {Karastergiou}, {Gupta}, {Johnston},
  {Bhat}, \& {Lyne}}]{Kramer2003}
{Kramer}, M., {Karastergiou}, A., {Gupta}, Y., {et~al.} 2003, \aap, 407, 655

\bibitem[{{Law} {et~al.}(2017){Law}, {Abruzzo}, {Bassa}, {Bower},
  {Burke-Spolaor}, {Butler}, {Cantwell}, {Carey}, {Chatterjee}, {Cordes},
  {Demorest}, {Dowell}, {Fender}, {Gourdji}, {Grainge}, {Hessels}, {Hickish},
  {Kaspi}, {Lazio}, {McLaughlin}, {Michilli}, {Mooley}, {Perrott}, {Ransom},
  {Razavi-Ghods}, {Rupen}, {Scaife}, {Scott}, {Scholz}, {Seymour}, {Spitler},
  {Stovall}, {Tendulkar}, {Titterington}, {Wharton}, \& {Williams}}]{Law2017}
{Law}, C.~J., {Abruzzo}, M.~W., {Bassa}, C.~G., {et~al.} 2017, \apj, 850, 76

\bibitem[{{Lawson} {et~al.}(1987){Lawson}, {Mayer}, {Osborne}, \&
  {Parkinson}}]{Lawson1987}
{Lawson}, K.~D., {Mayer}, C.~J., {Osborne}, J.~L., \& {Parkinson}, M.~L. 1987,
  \mnras, 225, 307

\bibitem[{{Li}(2006)}]{Li2006}
{Li}, X.-D. 2006, \apjl, 646, L139

\bibitem[{{Lorimer}(2011)}]{Lorimer2011}
{Lorimer}, D.~R. 2011, {SIGPROC: Pulsar Signal Processing Programs},
  Astrophysics Source Code Library, ascl:1107.016

\bibitem[{{Lorimer} {et~al.}(2007){Lorimer}, {Bailes}, {McLaughlin},
  {Narkevic}, \& {Crawford}}]{Lorimer2007}
{Lorimer}, D.~R., {Bailes}, M., {McLaughlin}, M.~A., {Narkevic}, D.~J., \&
  {Crawford}, F. 2007, Science, 318, 777

\bibitem[{{Lorimer} \& {Kramer}(2004)}]{HANDBOOK}
{Lorimer}, D.~R., \& {Kramer}, M. 2004, {Handbook of Pulsar Astronomy}

\bibitem[{{Lorimer} {et~al.}(1995){Lorimer}, {Yates}, {Lyne}, \&
  {Gould}}]{Lorimer1995}
{Lorimer}, D.~R., {Yates}, J.~A., {Lyne}, A.~G., \& {Gould}, D.~M. 1995,
  \mnras, 273, 411

\bibitem[{{Lyne} {et~al.}(2009){Lyne}, {McLaughlin}, {Keane}, {Kramer},
  {Espinoza}, {Stappers}, {Palliyaguru}, \& {Miller}}]{Lyne2009}
{Lyne}, A.~G., {McLaughlin}, M.~A., {Keane}, E.~F., {et~al.} 2009, \mnras, 400,
  1439

\bibitem[{{Lyubarsky}(2014)}]{Lyubarsky2014}
{Lyubarsky}, Y. 2014, \mnras, 442, L9

\bibitem[{{Lyutikov}(2002)}]{Lyutikov2002}
{Lyutikov}, M. 2002, \apjl, 580, L65

\bibitem[{{Manchester} {et~al.}(2001){Manchester}, {Lyne}, {Camilo}, {Bell},
  {Kaspi}, {D'Amico}, {McKay}, {Crawford}, {Stairs}, {Possenti}, {Kramer}, \&
  {Sheppard}}]{PMSURV2001}
{Manchester}, R.~N., {Lyne}, A.~G., {Camilo}, F., {et~al.} 2001, \mnras, 328,
  17

\bibitem[{{Maron} {et~al.}(2000){Maron}, {Kijak}, {Kramer}, \&
  {Wielebinski}}]{Maron2000}
{Maron}, O., {Kijak}, J., {Kramer}, M., \& {Wielebinski}, R. 2000, \aaps, 147,
  195

\bibitem[{{McLaughlin} {et~al.}(2006){McLaughlin}, {Lyne}, {Lorimer}, {Kramer},
  {Faulkner}, {Manchester}, {Cordes}, {Camilo}, {Possenti}, {Stairs}, {Hobbs},
  {D'Amico}, {Burgay}, \& {O'Brien}}]{McLaughlin2006}
{McLaughlin}, M.~A., {Lyne}, A.~G., {Lorimer}, D.~R., {et~al.} 2006, \nat, 439,
  817

\bibitem[{{McLaughlin} {et~al.}(2009){McLaughlin}, {Lyne}, {Keane}, {Kramer},
  {Miller}, {Lorimer}, {Manchester}, {Camilo}, \& {Stairs}}]{McLaughlin2009}
{McLaughlin}, M.~A., {Lyne}, A.~G., {Keane}, E.~F., {et~al.} 2009, \mnras, 400,
  1431

\bibitem[{{Meyers} {et~al.}(2017){Meyers}, {Tremblay}, {Bhat}, {Shannon},
  {Kirsten}, {Sokolowski}, {Tingay}, {Oronsaye}, \& {Ord}}]{Meyers2017}
{Meyers}, B.~W., {Tremblay}, S.~E., {Bhat}, N.~D.~R., {et~al.} 2017, ArXiv
  e-prints, arXiv:1709.03651

\bibitem[{{Michilli} {et~al.}(2018){Michilli}, {Seymour}, {Hessels}, {Spitler},
  {Gajjar}, {Archibald}, {Bower}, {Chatterjee}, {Cordes}, {Gourdji}, {Heald},
  {Kaspi}, {Law}, {Sobey}, {Adams}, {Bassa}, {Bogdanov}, {Brinkman},
  {Demorest}, {Fernandez}, {Hellbourg}, {Lazio}, {Lynch}, {Maddox}, {Marcote},
  {McLaughlin}, {Paragi}, {Ransom}, {Scholz}, {Siemion}, {Tendulkar}, {Van
  Rooy}, {Wharton}, \& {Whitlow}}]{Michilli2018}
{Michilli}, D., {Seymour}, A., {Hessels}, J.~W.~T., {et~al.} 2018, ArXiv
  e-prints, arXiv:1801.03965

\bibitem[{{Mickaliger} {et~al.}(2018){Mickaliger}, {McEwen}, {McLaughlin}, \&
  {Lorimer}}]{Mickaliger2018}
{Mickaliger}, M.~B., {McEwen}, A.~E., {McLaughlin}, M.~A., \& {Lorimer}, D.~R.
  2018, ArXiv e-prints, arXiv:1807.00143

\bibitem[{{Mickaliger} {et~al.}(2012){Mickaliger}, {McLaughlin}, {Lorimer},
  {Langston}, {Bilous}, {Kondratiev}, {Lyutikov}, {Ransom}, \&
  {Palliyaguru}}]{Mickaliger2012}
{Mickaliger}, M.~B., {McLaughlin}, M.~A., {Lorimer}, D.~R., {et~al.} 2012,
  \apj, 760, 64

\bibitem[{{Mikami} {et~al.}(2016){Mikami}, {Asano}, {Tanaka}, {Kisaka},
  {Sekido}, {Takefuji}, {Takeuchi}, {Misawa}, {Tsuchiya}, {Kita}, {Yonekura},
  \& {Terasawa}}]{Mikami2016}
{Mikami}, R., {Asano}, K., {Tanaka}, S.~J., {et~al.} 2016, \apj, 832, 212

\bibitem[{{Oppermann} {et~al.}(2018){Oppermann}, {Yu}, \&
  {Pen}}]{Oppermann2018}
{Oppermann}, N., {Yu}, H.-R., \& {Pen}, U.-L. 2018, \mnras, 475, 5109

\bibitem[{{Palliyaguru} {et~al.}(2011){Palliyaguru}, {McLaughlin}, {Keane},
  {Kramer}, {Lyne}, {Lorimer}, {Manchester}, {Camilo}, \&
  {Stairs}}]{Palliyaguru2011}
{Palliyaguru}, N.~T., {McLaughlin}, M.~A., {Keane}, E.~F., {et~al.} 2011,
  \mnras, 417, 1871

\bibitem[{{Pennucci} {et~al.}(2015){Pennucci}, {Possenti}, {Esposito}, {Rea},
  {Haggard}, {Baganoff}, {Burgay}, {Coti Zelati}, {Israel}, \&
  {Minter}}]{Pennucci2015}
{Pennucci}, T.~T., {Possenti}, A., {Esposito}, P., {et~al.} 2015, \apj, 808, 81

\bibitem[{{Petroff} {et~al.}(2016){Petroff}, {Barr}, {Jameson}, {Keane},
  {Bailes}, {Kramer}, {Morello}, {Tabbara}, \& {van Straten}}]{Petroff2016}
{Petroff}, E., {Barr}, E.~D., {Jameson}, A., {et~al.} 2016, PASA, 33, e045

\bibitem[{{Popov} \& {Stappers}(2007)}]{Popov2007}
{Popov}, M.~V., \& {Stappers}, B. 2007, \aap, 470, 1003

\bibitem[{{Rane} \& {Loeb}(2016)}]{Rane2016}
{Rane}, A., \& {Loeb}, A. 2016, ArXiv e-prints, arXiv:1608.06952

\bibitem[{{Rasmussen}(2006)}]{GPRBook}
{Rasmussen}, C.~E., W. C.~K.~I. 2006, Gaussian Processes for Machine Learning,
  1st edn. (The MIT Press)

\bibitem[{{Romero} {et~al.}(2016){Romero}, {del Valle}, \&
  {Vieyro}}]{Romero2016}
{Romero}, G.~E., {del Valle}, M.~V., \& {Vieyro}, F.~L. 2016, \prd, 93, 023001

\bibitem[{{Rutledge}(2006)}]{Rutledge2006}
{Rutledge}, R.~E. 2006, ArXiv Astrophysics e-prints, astro-ph/0609200

\bibitem[{{Spitler} {et~al.}(2016){Spitler}, {Scholz}, {Hessels}, {Bogdanov},
  {Brazier}, {Camilo}, {Chatterjee}, {Cordes}, {Crawford}, {Deneva}, {Ferdman},
  {Freire}, {Kaspi}, {Lazarus}, {Lynch}, {Madsen}, {McLaughlin}, {Patel},
  {Ransom}, {Seymour}, {Stairs}, {Stappers}, {van Leeuwen}, \&
  {Zhu}}]{Spitler2016}
{Spitler}, L.~G., {Scholz}, P., {Hessels}, J.~W.~T., {et~al.} 2016, \nat, 531,
  202

\bibitem[{{Spitler} {et~al.}(2018){Spitler}, {Herrmann}, {Bower}, {Chatterjee},
  {Cordes}, {Hessels}, {Kramer}, {Michilli}, {Scholz}, {Seymour}, \&
  {Siemion}}]{Spitler2018}
{Spitler}, L.~G., {Herrmann}, W., {Bower}, G.~C., {et~al.} 2018, ArXiv
  e-prints, arXiv:1807.03722

\bibitem[{{Vieyro} {et~al.}(2017){Vieyro}, {Romero}, {Bosch-Ramon}, {Marcote},
  \& {del Valle}}]{Vieyro2017}
{Vieyro}, F.~L., {Romero}, G.~E., {Bosch-Ramon}, V., {Marcote}, B., \& {del
  Valle}, M.~V. 2017, \aap, 602, A64

\bibitem[{{Wang} {et~al.}(2007){Wang}, {Manchester}, \& {Johnston}}]{Wang2007}
{Wang}, N., {Manchester}, R.~N., \& {Johnston}, S. 2007, \mnras, 377, 1383

\bibitem[{{Waxman}(2017)}]{Waxman2017}
{Waxman}, E. 2017, \apj, 842, 34

\bibitem[{{Yao} {et~al.}(2017){Yao}, {Manchester}, \& {Wang}}]{YMW2017}
{Yao}, J.~M., {Manchester}, R.~N., \& {Wang}, N. 2017, \apj, 835, 29

\end{thebibliography}

\end{document}